\documentclass[11pt]{JHEP3}

\usepackage{amsmath,amssymb}
\usepackage[T1]{fontenc} 
\usepackage{slashed}
\newcommand{\be}{\begin{equation}}
\newcommand{\ee}{\end{equation}}
\newcommand{\ben}{\begin{eqnarray}}
\newcommand{\een}{\end{eqnarray}}
\newcommand{\nn}{\nonumber}

\newcommand{\ads}{{\text{AdS}_5}}

\newcommand{\refb}[1]{(\ref{#1})}

\def\one{{\hbox{ 1\kern-.8mm l}}}
\def\zero{{\hbox{ 0\kern-1.5mm 0}}}

\def\cn{{\mathcal N}}

\def\tr{{\rm Tr}}

\def\l{\left}
\def\r{\right}

\def\f{\phi}

\def\s{\sigma}
\def\g{\gamma}
\def\G{\Gamma}

\def\o[#1]{{\rm O}\left({#1}\right)}
\def\dotl[#1,#2]{\left\langle #1, #2 \right\rangle}
\def\dotlb[#1,#2]{[ #1, #2 ]}
\def\dotp[#1,#2]{(#1) \cdot (#2)}
\def\>{\rangle}
\def\<{\langle}

\def\ss2{\text{S}^2}
\def\sn{\ss2/\mathbb{Z}_N}
\def\adss{\text{AdS}_2}
\def\adsn{\text{AdS}_2/\mathbb{Z}_N}
\def\s2s2{\ss2\otimes\ss2}
\def\ads2s2{\adss\otimes\ss2}
\def\s2s2n{\l(\ss2\otimes\ss2\r)/\mathbb{Z}_N}
\def\ads2s2n{\l(\adss\otimes\ss2\r)/\mathbb{Z}_N}
\def\zn{\mathbb{Z}_N}
\def\AAA{\mathcal{A}}

\title{Heat Kernels on the $\adss$ cone and Logarithmic Corrections to Extremal Black Hole Entropy}

\author{Rajesh Kumar Gupta$^a$\footnote{ rgupta AT ictp DOT it}, Shailesh Lal$^b$\footnote{shailesh DOT hri AT gmail DOT com}, Somyadip Thakur$^c$\footnote{somyadip AT cts DOT iisc DOT ernet DOT in}\\

$^a$ $\,$ICTP, High Energy, Cosmology and Astroparticle Physics,\\
$\;$ $\,$Strada Costiera 11, 34151, Trieste, Italy.\\

$^b$ $\,$Department of Physics and Astronomy, \\
$\;$ $\,$Seoul National University \\
$\;$ $\,$Seoul 151-747, Korea \\

$^c$ $\,$Centre for High Energy Physics,\\
$\;$ $\,$Indian Institute of Science, C.V. Raman Avenue,\\
$\;$ $\,$Bangalore 560012, India.\\
}

\abstract{We develop new techniques to efficiently evaluate heat kernel coefficients for the Laplacian in the short-time expansion on spheres and hyperboloids with conical singularities. We then apply these techniques to explicitly compute the logarithmic contribution to black hole entropy from an $\cn=4$ vector multiplet about a $\zn$ orbifold of the near-horizon geometry of quarter--BPS black holes in $\cn=4$ supergravity. We find that this vanishes, matching perfectly with the prediction from the microstate counting. We also discuss possible generalisations of our heat kernel results to higher-spin fields over $\zn$ orbifolds of higher-dimensional spheres and hyperboloids.}

\preprint{SNUTP13-006}
\begin{document} 

\section{Introduction}
One of the main successes of string theory as a theory of quantum gravity has been to provide a statistical interpretation of black hole entropy \cite{Strominger:1996sh}. This program has been very explicitly carried out for a class of supersymmetric black holes and a detailed matching of the string answer has been done in the large charge limit, where the string answer explicitly matches with the Bekenstein--Hawking, or more generally, the Wald entropy \cite{Wald:1993nt} of the black hole in question \cite{Dijkgraaf:1996it,LopesCardoso:2004xf,Shih:2005uc,Jatkar:2005bh, Dabholkar:2006xa,David:2006ji,David:2006ru,David:2006yn}. We refer the reader to \cite{Sen:2007qy} for a review and a more exhaustive set of references\footnote{This matching extends to the negative discriminant states in $\cn=4$ string theories as well, provided the macroscopic configurations are carefully identified \cite{Dabholkar:2007vk,Sen:2011mh,Chowdhury:2012jq}.}. However, the string answer is the full quantum answer for the entropy of the black hole and therefore also contains corrections to the Wald formula and it is interesting to ask if a direct physical interpretation of these corrections is possible, at least in some cases. In this paper, we will explore this question in the specific context of exponentially suppressed corrections to the microscopic degeneracy \cite{Banerjee:2008ky}. 

In the context of extremal black holes there is a parallel way of posing the black hole entropy problem. In particular, the near horizon geometry of such black holes is always of the form $\adss\otimes K$ where $K$ is a compact manifold. Then, it was proposed in \cite{Sen:2008vm} that the full quantum answer for the microscopic degeneracy associated with the black hole horizon is contained in the quantum entropy function, defined as the string path integral over all spacetimes which asymptote to the black hole near horizon geometry. The finite part of this path integral is expected to contain the entropy of the black hole at the quantum level. In particular, for extremal black holes carrying charges $\vec{q}\equiv q_i$ \cite{Sen:2008vm},
\begin{equation}\label{qef}
d_{hor}\l(\vec{q}\r)\equiv\l\langle\exp\l[i\oint q_i d\theta\mathcal{A}_\theta^i\r] \r\rangle_{\adss}^{finite},
\end{equation}
where $d_{hor}$ is the full quantum degeneracy associated with the black hole horizon, and $\mathcal{A}_\theta^i$ is the component of the $i^{\text{th}}$ gauge field along the boundary of the $\adss$. In this picture the entropy associated to the horizon degrees of freedom of an extremal blackhole is the free energy corresponding to the partition function \eqref{qef}. This proposal has been tested in a variety of ways, for which we refer the reader to \cite{Sen:2009vz,Sen:2009md,Sen:2010ts,Banerjee:2008ky,Banerjee:2009uk,Murthy:2009dq,Sen:2009gy} and more generally the lectures \cite{Mandal:2010cj} for an overview. A particularly non-trivial test of this proposal is that the leading quantum corrections in the large charge limit, which scale as $\log\l(\text{charges}\r)$, to the semi-classical Bekenstein-Hawking formula as predicted from the string computation can be reproduced from the quantum entropy function for $\cn=4$ and $\cn=8$ string theory \cite{Banerjee:2010qc,Banerjee:2011jp}. This is obtained by expanding the quantum entropy function about the saddle-point defined by the near-
horizon geometry of the black hole.

At this point it is natural to ask if the exponentially suppressed contributions to the microscopic degeneracy have an interpretation from the point of view of the quantum entropy function. It is proposed that these should be thought of as saddle-points of the string path integral of \eqref{qef} where the near-horizon geometry of the black hole is quotiented by a $\zn$ orbifold which we shall shortly review and define. This proposal already passes two non-trivial tests. Firstly, it may be shown using appropriate coordinate transformations that this configuration has the appropriate asymptotic behaviour for it to be considered an admissible saddle-point to \eqref{qef}, and secondly, it gives the correct leading term for $d\l(Q,P\r)$ in the large-charge limit \cite{Sen:2009vz}.

In this paper, we shall initiate a program aimed at providing a quantum test of this proposal. The overall goal is to compute the log correction in the $\zn$ orbifold of the attractor geometry and match it against the microscopic answer available from the string computation. We do not solve this problem in its entirety here. This paper is devoted to developing firstly the necessary techniques that will permit us to solve this problem efficiently, and secondly to carrying out a natural first consistency check of this conjecture as an application of these techniques. In particular, as we shall review soon, the log correction about these exponentially suppressed terms actually vanishes in $\cn=4$ string theory for arbitrary values of $N$. As a consistency check against this microscopic result, we shall explicitly demonstrate that the contribution of a single $\cn=4$ vector multiplet to the log term vanishes about this saddle point. The full gravity calculation is technically more involved, and is in progress.

A brief overview of this paper is as follows. In Section \ref{microrev} we briefly review the string theory answer for the entropy of quarter--BPS black holes and the quantum entropy function proposal for their macroscopic dual. These geometries are essentially $\zn$ orbifolds of the near horizon geometry of the black hole in question. Section \ref{3} is a brief review of the heat kernel method as applied to extract logarithmic terms from the partition function of a quantum theory. As the overall goal of the paper is to compute the heat kernel about these orbifolds and extract out log corrections to the black hole entropy, Sections \ref{4} and \ref{lapkernelads2s2n} develop new techniques for computing the heat kernel in these $\zn$ orbifolds of $\adss$, $\ss2$ and $\adss\otimes\ss2$. The AdS results are reached by analytic continuation from the sphere, for the validity of which some evidence is provided in Section \ref{4}. Finally in Section \ref{graviphoton} we compute the contribution to the log term from 
a single $\cn=4$ vector multiplet. We find that though individual contributions from gauge fields, scalars and fermions have very nontrivial $N$ dependence, the final log contribution still vanishes when zero modes are properly accounted for. This matches perfectly with our expectations from the microscopic results reviewed in Section \ref{microrev}. We then conclude.
\section{Quarter-BPS Black Holes in $\cn=4$ String Theory and their Entropy}\label{microrev}
We begin with a brief review of black hole entropy for $\cn=4$ string theories. In particular, we shall discuss the microscopic formula for black hole entropy obtained from string theory, and the Quantum Entropy Function formalism \cite{Sen:2008vm} which is a prescription for computing the full quantum entropy associated with the black hole horizon. We then propose a quantum test of the proposal of \cite{Sen:2008vm} analogous to the tests performed in \cite{Banerjee:2010qc,Banerjee:2011jp}.
\subsection{Exponentially Suppressed Corrections to Black Hole Entropy}\label{micro}
In this section we briefly review the stringy origin of the entropy of ${1\over 4}$--BPS black holes in $\cn=4$ string theories \cite{Dijkgraaf:1996it,LopesCardoso:2004xf,Jatkar:2005bh,David:2006yn}. We refer the reader to \cite{Sen:2007qy} for an indepth account and a more extensive set of references. In particular, we will study the degeneracy of ${1\over 4}$--BPS dyons in the large charge limit in which the results can be compared to the gravity side. This is essentially a review of the results of \cite{Banerjee:2008ky}. We shall concentrate especially on the $\cn=4$ string theory obtained by compactifying Type IIB string theory on $K3\times T^2$ or heterotic string theory on $T^6$. In this case the electric charge $Q$, and magnetic charge $P$ of a dyon are 28--dimensional vectors and the dyon degeneracy is given in terms of the Igusa cusp form $\Phi_{10}$ by
\be\label{deg}
d(Q,P)=(-1)^{Q\cdot P+1}\int_{{\mathcal C}}d\hat\rho d\hat\sigma d\hat v e^{-\pi i(\hat \rho P^2+\hat \sigma Q^2+2\hat v Q\cdot P)}\frac{1}{\Phi_{10}(\hat\rho,\hat\sigma,\hat v)}
\ee
where $\hat\rho$, $\hat\sigma$ and $\hat v$ are complex variables, and the integral is over $\mathcal{C}$, which is a three real-dimensional subspace of the $\mathbb{C}_3$ spanned by $\l(\hat{\rho},\hat{\sigma},\hat{v}\r)$. For more details, we refer the reader to \cite{Banerjee:2008ky,David:2006yn,Sen:2007qy}.
The poles of the integrand are the zeros of $\Phi_{10}$ which are located at 
\be\label{pol.1}
n_2(\hat\sigma\hat\rho-\hat v^2)+j\hat v+n_1\hat\sigma-\hat\rho m_1+m_2=0,\\
\ee 
where
\be
m_1,n_1,m_2,n_2\in {\mathbb Z},\quad j\in 2{\mathbb Z}+1,\quad m_1n_1+m_2n_2+\frac{j^2}{4}=\frac{1}{4}.
\ee
We note here that the equations \eqref{pol.1} are symmetric under the transformation $\l(\vec{m},\vec{n},j\r)\rightarrow\l(-\vec{m},-\vec{n}-j\r)$. We use this symmetry to set $n_2\geq 0$. In the remainder of this paper, we shall focus on the case of $n_2\geq 1$\footnote{The case of $n_2=0$ is also physically relevant. These poles capture the jumps in degeneracy as one moves across walls of marginal stability \cite{Sen:2007vb}.}. Further, for a given $n_2$, we can use symmetries of $\Phi_{10}$ to restrict the range of $m_1,n_1$ and $j$ to
\begin{equation}
0\leq n_1\leq n_2-1,\quad 0\leq m_1\leq n_2-1,\quad 0\leq j\leq 2n_2-1.
\end{equation}
We shall focus on how these results are used to evaluate the statistical entropy $S_{stat}\l(Q,P\r)\equiv \ln d\l(Q,P\r)$ and extract its behaviour in the large charge limit. The strategy one follows in evaluating the integral in (\ref{deg}) has been explicated in \cite{Sen:2007qy}. The procedure is to deform the contour $\mathcal{C}$ to values of $(\hat\rho_2,\hat\sigma_2,\hat v_2)$ of the order of 1/charge. In that case, the integral in \eqref{deg} receives contributions from the deformed contour and from the poles of the integrand which were crossed during this deformation. It can further be shown \cite{Sen:2007qy} that the dominant contribution to $d\l(Q,P\r)$ arises from the the poles \eqref{pol.1} of the integrand, while the contribution from the deformed contour is subleading. Then using the residue theorem one first reduces the integration variables in \eqref{deg} from three to two by performing the integration over $\hat v$. The integrations over $\hat{\rho}$ and $\hat{\sigma}$ are performed using 
the method of steepest descent. It may be shown that for a given choice of $\vec{m},\vec{n},j$, the saddle point lies at \cite{Dijkgraaf:1996it,Sen:2007qy}
\be\label{sadlpt}
(\hat\rho,\hat\sigma,-\hat v)=\frac{i}{2n_2\sqrt{Q^2P^2-(Q\cdot P)^2}}\left(Q^2,P^2,Q\cdot P\right)-\frac{1}{n_2}\left(n_1,-m_1,\frac{j}{2}\right)
\ee
Using the above results, it is finally found that the degeneracy at the saddle point is given by
\ben\label{deg2}
d(Q,P)_{n_2}&=&\frac{(-1)^{P\cdot Q}}{n_2}\exp\left(\pi\sqrt{Q^2P^2-(Q\cdot P)^2}/n_2\right)\exp\left[\frac{i\pi}{n_2}(n_1P^2-m_1Q^2+jQ\cdot P)\right]\nn\\&&\times\left[\det(C\hat\Omega+D)^{\kappa+2}g(\rho)^{-1}g(\sigma)^{-1}(1+\mathcal{O}(q^{-2}))\right]_{saddle},
\een
where
\be
g(\rho)=\eta(\rho)^{24},\quad q^2=(Q^2,P^2,Q\cdot P),
\ee
Where $\eta(\rho)$ is Dedekind eta function and $\rho$, $\sigma$ and $v$ are variables defined in terms of $\l(\hat{\rho},\hat{\sigma},\hat{v}\r)$ as
\be
\begin{pmatrix}
\rho & v\\
v & \sigma
\end{pmatrix}=\Omega=(A\hat\Omega+B)(C\hat\Omega+D)^{-1},\quad \hat\Omega=\begin{pmatrix}
\hat\rho & \hat v\\
\hat v & \hat\sigma
\end{pmatrix}
\ee
Here the matrices $A$, $B$, $C$ and $D$ are chosen such that 
\be
v=\frac{n_2(\hat\sigma\hat\rho-\hat v^2)+j\hat v+n_1\hat\sigma-m_1\hat\rho+m_2}{{\text det}(C\hat\Omega+D)}
\ee
Thus in the un-hatted variables the pole (\ref{pol.1}) is at $v=0$.
The exponential term, apart from phase factor, in the degeneracy gives the leading entropy which goes like $S_{BH}\over n_2$. Thus we see that dominant contribution to the entropy comes from the saddle point for which $n_2=1$. 
Now in order to see the log correction in the entropy and to compare with macroscopic entropy from gravity side, we take all the components of the charge vector to be very large. We then uniformly scale all charges as $Q\rightarrow \Lambda \tilde Q,\quad P\rightarrow\Lambda \tilde P$ where $\Lambda$ is very large number. In this case the leading contribution to entropy and hence area scales as
\be
\pi\sqrt{Q^2P^2-(Q\cdot P)^2}\rightarrow \Lambda^2\pi\sqrt{\tilde Q^2\tilde P^2-(\tilde Q\cdot \tilde P)^2}
\ee
Thus the terms in the degeneracy (\ref{deg2}) which are proportional to $\Lambda$ will give logarithm correction to entropy. It is easy to see that  $(\hat\rho,\hat\sigma,-\hat v)$ or  $\hat\Omega$ at the saddle point do not depend on $\Lambda$ and hence the matrices A, B, C, D and $\Omega|_{saddle}$ are independent of $\Lambda$. Thus we see that the term in the square bracket of $d(Q,P)_{n_2}$ goes as $\mathcal{O}(1)+\mathcal{O}(\frac{1}{\Lambda^2})$. Thus the entropy from the saddle point is given by
\ben
S_{n_2}&&=\ln d(Q,P)_{n_2}=\left(\pi\sqrt{Q^2P^2-(Q\cdot P)^2}/n_2\right)+\ln[\mathcal{O}(1)+\mathcal{O}(\frac{1}{\Lambda^2})]\nn\\&&\sim\left(\pi\sqrt{Q^2P^2-(Q\cdot P)^2}/n_2\right)+\frac{\mathcal{O}(1)}{\Lambda^2}+.......
\een
Thus in this example of $\cn=4$ string theory there is no log corrections to the leading exponential term. This statement is independent of the value of $n_2$. In Section \ref{graviphoton} we will show from the quantum entropy function that the log correction for a single $\cn=4$ vector multiplet vanishes. Firstly, this is entirely consistent with the above result, and secondly it gives evidence that there should be no log corrections for any $\cn=4$ string theory at any value of $n_2$. To prove this statement explicitly, we will need to include the contribution of the full gravity multiplet, which is work in progress.\\
\subsection{The Quantum Entropy Function and their Macroscopic Origin}\label{macro}
We have seen in Section \ref{micro} that while the statistical entropy $d\l(Q,P\r)$ grows as $e^{\pi\sqrt{\Delta}}$ in the large charge limit, it also has exponentially suppressed contributions which grow as $e^{{\pi\sqrt{\Delta}\over n_2}}$ in the same scaling limit, where $n_2>1$. In the context of extremal black holes there is a complementary formula for the full quantum entropy of the black hole, known as the quantum entropy function \cite{Sen:2008vm}. The proposal is that the degeneracy associated with the horizon degrees of freedom of the extremal black hole is captured by the string partition function over spacetimes that asymptote to the near-horizon geometry of the black hole\footnote{The near-horizon geometry of an extremal black hole is $\adss$ times a compact manifold $K$. Due to the presence of the $\adss$ factor, the string path integral diverges. There is however a well-defined prescription by which a finite part of this partition function may be extracted \cite{Sen:2008vm}. By now, there is 
very non-trivial evidence that this finite part does indeed correctly capture the degeneracy in the horizon degrees of freedom of the black hole, see \cite{Mandal:2010cj} for a review.}. In particular, the dominant contribution, which corresponds to the Bekenstein-Hawking (more generally, Wald) entropy corresponds to the following saddle-point of the string path integral\footnote{We write the solution in the supergravity obtained by compactifying Type IIB Supergravity on $K3$.}
\begin{equation}\label{leadingsaddle}
\begin{split}
ds^2&= v\l(d\eta^2+\sinh^2\eta d\theta^2\r)+u\l(d\psi^2+\sin^2\psi d\phi^2\r) +{R^2\over\tau_2}\vert dx_4+\tau dx_5\vert^2,\\
G^I&= {1\over 8\pi^2}\l[Q_I\sin\psi dx^5\wedge d\psi\wedge d\phi+P_I\sin\psi dx^4\wedge d\psi\wedge d\phi +\text{dual}\r],\\
V_I^i&=\text{constant},\quad V_I^r=\text{constant}.
\end{split}
\end{equation}
We refer the reader to \cite{Banerjee:2009af} for more details about this solution. Here we merely mention that the parametres $u,v,R,\tau$ and $V$ are determined completely in terms of the electric and magnetic charges $\l(Q,P\r)$ of the black hole, and $u=v$. $R$ and $\tau$ do not scale with the $AdS$ radius $u$.
A particularly non-trivial test of this proposal is that expanding in quadratic fluctuations about this saddle point yields a term proportional to $\log a$, where $a$ is the radius of the $\adss$ and $\ss2$ submanifolds. This matches precisely with the $\log a$ term extracted from the microscopic degeneracy as in section \ref{micro}, in that both of them are zero\footnote{The corresponding computation for black holes in $\cn=8$ string theory yields a non-vanishing quantity which is also found to match with the microscopic formula \cite{Banerjee:2011jp}.} \cite{Banerjee:2010qc,Banerjee:2011jp}. At this point, it is natural to ask if the exponentially suppressed corrections to the microscopic degeneracy have a proposed counterpart in the quantum entropy function formalism, and it turns out that the answer is yes. These correspond to the following orbifolds of the geometry \eqref{leadingsaddle} \cite{Banerjee:2008ky,Sen:2009vz}
\begin{equation}\label{horizonorbifold}
\theta\mapsto\theta+{2\pi\over N},\quad\phi\mapsto\phi-{2\pi\over N},\quad x^5\mapsto x^5+{2\pi\kappa\over N},\quad \text{gcd}\l(\kappa,N\r)=1.
\end{equation}
It may be seen, for one, that the leading contribution to the entropy from these saddle points is ${A\over 4N}$ \cite{Banerjee:2008ky,Sen:2009vz}, which precisely matches the leading contribution of the exponentially suppressed terms with the identification $n_2 \leftrightarrow N$. Additionally, it can also be shown \cite{Sen:2009vz} that these geometries obey the appropriate fall-off conditions for them to be included in the string path integral in \eqref{qef}. We refer the reader to \cite{Banerjee:2008ky,Sen:2009vz,Banerjee:2009af} for more details regarding this orbifold. In this paper, we will perform a further test of this proposal. In particular, as we have seen above, there exist a family of $\cn=4$ string theories with an arbitrary number of vector multiplets for which the term proportional to $\log a$ vanishes even about these exponentially suppressed terms. It must therefore be that the contribution of a single $\cn=4$ vector multiplet to the log term vanishes. In this paper, we will set up heat kernel techniques to tackle this problem, and explicitly verify this conjecture. In particular, we will revisit the computation of the one-loop partition function carried out in the near 
horizon $\adss\otimes\ss2$ geometry of a ${1\over 4}$--BPS black hole in \cite{Banerjee:2010qc} and carry out the same computation on the $\zn$ orbifold of the geometry where the $\zn$ quotients the $\theta$ and $\phi$ angles of $\adss$ and $\ss2$ as in \eqref{horizonorbifold}. We will then demonstrate that the contribution of the $\cn=4$ vector multiplet in this orbifolded background indeed vanishes, in accordance with the expectation from the microscopic results\footnote{The reader might worry that by computing only over the spectrum of massless fields in four dimensions one is missing the $N$ dependence which would come from the $x^5$ quotient. This is true. However, we are looking to compute the contribution to the partition function which scales non-trivially with $a$ at large $a$. This scale appears only in the $\adss\otimes\ss2$ part of the near-horizon geometry. Hence, for this purpose, it is sufficient to reduce onto the four-dimensional part of the attractor geometry and, as argued in \cite{
Banerjee:2010qc,Sen:2012dw} only the massless modes in this effective geometry will contribute to the log term.}.
\section{The Heat Kernel Method and Logarithmic Corrections to the Partition Function}\label{3}
This section is a brief review of the heat kernel method as applied to extract the logarithmic corrections to the partition function of a generic quantum field theory. In particular, we shall focus on an effective field theory in $d+1$ dimensions, defined on a background with length scale $a$, and show how the term that scales as $\log a$ may be extracted for large $a$ from the heat kernel. We refer the reader to \cite{Banerjee:2010qc}--\cite{Bhattacharyya:2012ye} where more details and references may be found. Firstly, it may be shown that the term scaling as $\log a$ is sensitive only to the two-derivative sector of the theory, and further only receives contributions from massless fields, and at one-loop only \cite{Sen:2012dw}. We will therefore concentrate on the one-loop partition function of the theory. We consider, therefore, the partition function 
\begin{equation}
\mathcal{Z}\l[\Phi\r]=\int\l[\mathcal{D}\Phi\r] e^{-{1\over\hbar}S\l[\Phi\r]}.
\end{equation}
In the limit $\hbar\rightarrow 0$, this is dominated by classical configurations $\Phi_{cl}$, which solve
\begin{equation}
{\delta\over\delta\Phi}S\mid_{\Phi=\Phi_{cl}}=0.
\end{equation}
We decompose the field $\Phi$ into small fluctuations about the configuration $\Phi_{cl}$
\begin{equation}
\Phi=\Phi_{cl}+\phi,
\end{equation}
in which case
\begin{equation}
S\l[\Phi\r]\simeq S\l[\Phi_{cl}\r]+\int d^{d+1}x\sqrt{g}\phi\l(x\r)D\phi\l(x\r),
\end{equation}
where $D$ is the kinetic operator for the fluctuation $\phi$. Higher-order terms in $\phi$ correspond to higher-loop corrections and have been omitted here. The one-loop partition function is then given by
\begin{equation}
\mathcal{Z}_{1-\ell}={\det}^{-{1\over 2}}\l(D\r).
\end{equation}
To define the determinant, we use the identity
\begin{equation}
\det D=\prod_n \kappa_n,
\end{equation}
where the $\kappa_n$ are the eigenvalues of $D$. In that case, it is easy to see that
\begin{equation}\label{logdet}
-\log\det D =-\sum_n \log\kappa_n =\sum_n\int\sqrt{g}d^{d+1}x\, \int_0^\infty {dt\over t}e^{-t\kappa_n}\Psi_n^*\l(x\r)\Psi_n\l(x\r),
\end{equation}
where the $\Psi_n$ are the normalised eigenfunctions of the operator $D$ belonging to the eigenvalue $\kappa_n$.Though this expression has been written for a discrete non-degenerate spectrum, the generalisation to the continuous and degenerate cases is apparent. The final expression in \eqref{logdet} is just the trace of the heat kernel for the operator $D$
\begin{equation}
K_{ab}\l(x,y;t\r)=\sum_{n}\Psi_{n,a}\l(x\r)\Psi^*_{n,b}\l(y\r) e^{-t \kappa_n},
\end{equation}
i.e.
\begin{equation}\label{nonzero}
-\log\det D = \int_\epsilon^\infty {dt\over t} K\l(t\r),
\end{equation}
where
\begin{equation}
K\l(t\r)=\sum_n\int\sqrt{g}d^{d+1}x\, e^{-t\kappa_n}\Psi_n^*\l(x\r)\Psi_n\l(x\r),
\end{equation}
and $\epsilon$ is a UV cutoff. In the above analysis we have assumed that all the $\kappa_n$ are positive definite. This is indeed true for the Laplacian on spheres, except for the constant scalar mode, but the Laplacian on hyperboloids is only positive semi-definite. In particular, there exist fields on hyperboloids for which the Laplacian has non-trivial zero modes. This includes one-forms on $\adss$ \cite{Camporesi:1994ga}, and more generally $p$-forms on AdS$_{2p}$ \cite{Camporesi}.\footnote{For these fields, by `Laplacian', we mean the Hodge Laplacian, given by $\Delta=d\delta+\delta d$, where $\delta$ is the adjoint of the exterior derivative, defined with respect to the background metric by the Hodge star.} To illustrate how the above analysis is modified in the presence of zero modes, consider the finite-dimensional Gaussian integral
\begin{equation}
\mathcal{Z}=\int \prod_{i=1}^n dx_i \, e^{-\sum_{i,j=1}^n x^i M_{ij} x^j},
\end{equation}
where $M$ has already been diagonalised. In particular, $M=diag\l(\lambda_1,\ldots,\lambda_{n-1},0\r)$. In that case, it is easy to see that $\mathcal{Z}$ is given by
\begin{equation}
\mathcal{Z}=\l({\det}' M\r)^{-{1\over 2}}\int dx_n,
\end{equation}
where $\det'\l(M\r)=\prod_{i=1}^{n-1}\kappa_i$. Hence, the path integral decomposes into the determinant of the operator $M$ evaluated over the subspace of its non-zero eigenvectors, with a residual zero-mode integral remaining. This is precisely what happens for the path integral as well. We find that
\begin{equation}
\mathcal{Z}_{1-\ell}={{\det}'D}^{-{1\over 2}}\cdot\mathcal{Z}_{\text{zero}},
\end{equation}
where $\det'D$ is the determinant of $D$ evaluated over its non-zero modes, and $\mathcal{Z}_{\text{zero}}$ is the residual zero-mode integral. We will therefore compute the determinant of the Laplacian explicitly over non-zero modes only, and separately analyse the zero mode integral.\footnote{This is in slight contrast to the methods of \cite{Banerjee:2010qc}--\cite{Bhattacharyya:2012ye}, where the heat kernel was computed over the full set of eigenmodes of the fields, including zero modes, and then the zero mode contribution was subtracted out to give the determinant over non-zero modes. The two methods are of course equivalent.} and In order to extract the zero mode contribution to the log term, it is sufficient to determine how $\mathcal{Z}_{\text{zero}}$ scales with $a$. This has been done in \cite{Banerjee:2010qc} and we do not repeat it here. The final result is that if the operator $D$ has $n_0$ zero modes when acting over a vector field, then \footnote{The more general result is that if an operator $D$ has $n_0$ zero modes and integration over a zero mode gives a factor $a^\beta$ then $\log \mathcal{Z}_{\text{zero}}= n_0\beta\log a +\mathcal{O}\l(1\r)$ \cite{Banerjee:2011jp,Sen:2012dw}.} 
\begin{equation}
\log \mathcal{Z}_{\text{zero}}= n_0\log a +\mathcal{O}\l(1\r).
\end{equation}
Now, we need to examine how the non-zero modes contribute to the log term. For this, we firstly observe that if $D$ is a Laplace-type operator defined on a background with a metric with an overall scale $a$, the eigenvalues of $D$ scale as ${1\over a^2}$. We then define a new variable $\bar{s}={t\over a^2}$ in terms of which \eqref{nonzero} becomes
\begin{equation}
-\log{\det}'D= \int_{\frac{\epsilon}{a^2}}^\infty {d\bar{s}\over\bar{s}} K\l(\bar{s}\r).
\end{equation}
We therefore find that
\begin{equation}
\log\mathcal{Z}_{\text{non-zero}}=-{1\over 2}\log{\det}'D= K_1\log a +\ldots,
\end{equation}
where the terms in $`\ldots$' do not scale as $\log a$ and $K_1$ is the order-1 term in the series expansion of $K\l(\bar{s}\r)$ about $\bar{s}=0$. We then find that the contribution to the free energy which scales as $\log a$ is given by
\begin{equation}\label{logcontr}
F_{\log} = \l(K_1+n_0\r)\log a.
\end{equation}
In the rest of this paper, we will compute the two contributions to the partition function for scalar, spin-half and vector fields on $\zn$ orbifolds of $\ss2$, $\adss$ and their product spaces. Finally, in section \ref{graviphoton} we will see that this contribution from a single $\cn=4$ vector multiplet vanishes in the $\zn$ orbifold of the near horizon geometry of a ${1\over 4}$--BPS black hole in $\cn=4$ supergravity, which precisely matches with our expectations from the microscopic computations of the entropy of this black hole.
\section{The Heat Kernel Laplacian on $\adss/\zn$.}\label{4}
In order to compute logarithmic corrections about these saddle-points from the Quantum Entropy Function, we need to evaluate the integrated heat kernel about the quotient spaces defined in Section \ref{macro}. As a warm-up, we will consider in this section the heat kernel of the scalar Laplacian on a $\zn$ orbifold of $\adss$ which we shortly specify. In principle, this answer can also be extracted by employing the Sommerfeld formula \cite{sommer1,sommer2,sommer3} on $\adss$.\footnote{See also \cite{Mann:1996bi,Mann:1996ze} for related work in AdS geometries, \cite{Fursaev:1996uz,DeNardo:1996kp} for related work on spherical geometries.} However, the answer is obtained in an integral form from which the $N$ dependence can only be extracted in a series expansion \cite{Mann}. We propose here an alternate method, which involved computing the integrated heat kernel on the corresponding $\zn$ orbifold of $\ss2$ by explicitly enumerating the spectrum and degeneracies of the Laplacian on this space, and performing an analytic continuation, which we motivate, to $\adss$. This has two benefits. Firstly, at a technical level, we obtain a new and efficient method for computing 
coefficients in the Seeley-de Wit expansion of the heat kernel on these conical spacetimes. This method can be easily employed, once set up, to extract coefficients to arbitrary power in the heat kernel time. In this manner, we will obtain the full $N$-dependence of the heat kernel coefficients to the required accuracy. Secondly, we will see that this strategy of computing the heat kernel by explicitly enumerating the spectrum and degeneracies of the quadratic operator will be of great use when we compute the heat kernel of the Laplacian on the higher-dimensional spaces $\ss2\otimes\ss2$, its $\zn$ orbifold, and when taking into account the effects of the graviphoton flux while computing the relevant one-loop determinants for the $\cn=4$ vector multiplet. Additionally, we note that the spectrum and degeneracies of the spin-$s$ Laplacian are very explicitly known on arbitrary-dimensional spheres and hyperboloids \cite{Camporesi:1994ga}. It is possible, though we do not carry out this program here, that our analysis may be extended to 
explicitly compute the heat kernel for higher-spin fields, in higher-dimensional spacetimes as well.
\subsection{The $\zn$ Orbifold of $\ss2$ and $\adss$}
We will consider quotients of the two-sphere, with metric given by
\begin{equation}\label{metrics2}
ds^2=a^2\l(d\rho^2+\sin^2\rho d\theta^2\r),
\end{equation}
and AdS$_2$ with metric 
\begin{equation}\label{metricads2}
ds^2=a^2\l(d\chi^2+\sinh^2\chi d\theta^2\r),
\end{equation}
which can be obtained from \refb{metrics2} by the analytic continuation $\rho\mapsto i\chi$, $a\mapsto ia$. The quotient we consider is
\begin{equation}
\theta\mapsto\theta+ {2\pi\over N}.
\end{equation}
Under this quotient, $\rho=0,\pi$ are fixed points on the sphere and $\chi=0$ is the fixed point on AdS$_2$. In the neighborhood of each fixed point (small enough that $\sin\rho\simeq\rho$ and $\sinh\chi\simeq\chi$), the corresponding geometries become
\begin{equation}
ds^2=dr^2+r^2 d\theta^2,
\end{equation}
where $r=a\rho$ or $a\chi$ as the case may be, i.e. in the neighborhood of the fixed point, the geometry is that of a cone. We have to evaluate the integrated heat kernel on these manifolds. The integrated heat kernel on $\adss$ and $\adsn$ suffers from a volume divergence which needs to be regulated carefully to capture the finite $N$-dependent terms in the heat kernel expansion. This is analogous to the problem encountered for the heat kernel on hyperboloids quotiented by freely acting orbifolds \cite{David:2009xg,Gopakumar:2011qs}. Even in those cases the heat kernel suffers from a volume divergence, but the regulated answer turns out to have an interesting dependence on the chemical potentials corresponding to the quotient. To obtain this $N$ dependence, we will adopt a strategy very similar to \cite{David:2009xg,Gopakumar:2011qs}; we will carry out the computation on the sphere, and analytically continue the result to the hyperboloid by a prescription we develop below.
\subsection{The Scalar Laplacian on $\sn$}\label{scalarsn}
We begin by recalling the well-known spectrum of the scalar Laplacian on $\ss2$. The eigenvalues of the Laplacian are given by
\begin{equation}
E_\ell = {\ell\l(\ell+1\r)\over a^2}, \quad \ell=0,1,2,\ldots,
\end{equation}
these eigenvalues are $d_\ell=\l(2\ell+1\r)$--fold degenerate, with the corresponding eigenfunctions being given by the spherical harmonics
\begin{equation}
Y_{\ell m}\l(\rho,\phi\r)= P_{\ell}^m\l(\cos\rho\r)e^{\pm im\phi}, \quad \vert m\vert \in \l[0,\ell\r], m \in \mathbb{Z}.
\end{equation}
The corresponding integrated coincident heat kernel is given by
\begin{equation}
I=\int d^2x\sqrt{g} K\l(x,x;t\r)=\sum_{\ell} \l(2\ell+1\r)e^{-{t\over a^2}\ell\l(\ell+1\r)}=\sum d_{\ell}e^{-tE_{\ell}}.
\end{equation}
We will now write down a corresponding expression for the heat kernel of the scalar Laplacian over the quotient space $\sn$, where the heat kernel is computed over scalar modes periodic under the $\zn$ quotient. Clearly, these are just the modes for which $m=Np$, where $p$ is an integer\footnote{This spectrum also includes the $\ell=0$ mode, which is a zero mode. As per the discussion in Section \ref{3}, we should perform the integral over the zero mode separately. However, using the methods of \cite{Banerjee:2010qc} one may show that the $\log a$ term extracted remains the same. For fields on $\ads2s2n$ which we encounter in Section \ref{lapkernelads2s2n} we shall indeed treat the zero mode integral separately.}. The expression for the integrated heat kernel for the scalar Laplacian on $\sn$ in terms of eigenvalues and degeneracies is then given by\footnote{It is easy to see that for a given $\ell=Np+q$, where $0\leq q<N$ is an integer, the allowed $m$ values are $\vert m\vert=0,N,2N,\ldots,Np$. The expression \refb{qsphereN} follows.}
\begin{equation}\label{qsphereN}
S= \sum_{p=0}^\infty {\l(2p+1\r)}\l[e^{-t{Np(Np+1)\over a^2}}+e^{-t{(Np+1)(Np+2)\over a^2}}+\ldots+e^{-t{(Np+N-1)(Np+N)\over a^2}}\r]\equiv \sum_{n=1}^N S_n,
\end{equation}
where 
\begin{equation}
S_n=\sum_{p=0}^\infty \l(2p+1\r)e^{-t{(Np+n-1)(Np+n)\over a^2}}.
\end{equation}
It turns out that the expression \refb{qsphereN} can be related to the integrated heat kernel over the unquotiented sphere $\ss2$, with some corrections which can be precisely computed. The final result should not be surprising; if we excise the fixed points on $\sn$, which house curvature singularities, the remaining space is completely smooth, and the heat kernel asymptotics can be computed from the Seeley-de Wit expansion where the coefficients are well-known \cite{Vassilevich:2003xt}. Intuitively, it should be possible to think of the final answer for the heat kernel on these singular spaces as arising from the smooth part of the manifold, with an additional contribution from the conical singularities \cite{Fursaev:1993qk}. We do indeed recover this structure in our final answer for $\sn$, which is also useful for fixing the analytic continuation to $\adsn$. 
\subsubsection{The Large Radius Approximation}\label{largerad}
As the subsequent computations are rather formal, it is useful to consider first the heat kernel on $\sn$ in a limit where the radius $a$ is very large\footnote{We would like to thank Justin David for suggesting this approach to us.}. We will focus on the dominant term in the large-$a$ expansion. In the next section, we make no such approximation, and write down the corresponding exact expression. Essentially, in the large radius limit, the sum over $p$ in \refb{qsphereN} can be replaced by an integral.In particular, we use the approximation that
\begin{equation}\label{intapprox}
\int_a^b f(x)\, dx \simeq {1\over 2}f(a)+f(a+1)+\ldots +f(b-1)+{1\over 2} f(b).
\end{equation}
The error associated with making this approximation is given by
\begin{equation}\label{interr}
E=\sum_{k=2}^\infty {B_k\over k!}\l(f^{\l(k-1\r)}\l(b\r)-f^{\l(k-1\r)}\l(a\r)\r).
\end{equation}
We will show that the error involved in making the approximation \refb{intapprox} is subleading in the large-$a$ expansion. Further, in this limit, the integrated heat kernel over the $\zn$ orbifold of $\ss2$ is just given by ${1\over N}$ times the integrated heat kernel over $\ss2$. For definiteness, we shall fix $N$ to 2; the computation for arbitrary $N$ is entirely analogous. In this case, the integrated heat kernel is given by the sum
\begin{equation}\label{qsphere}
S =\sum_{m=0}^\infty {\l(2m+1\r)}\l[e^{-{t\over a^2}2m(2m+1)}+e^{-{t\over a^2}(2m+1)(2m+2)}\r]\equiv S_1+S_2.
\end{equation}
Now, using \refb{intapprox}, $S_1$ and $S_2$ may be approximated as
\begin{equation}
S_1\simeq {1\over 2}+\int_0^\infty dm\l(2m+1\r)e^{-{t\over a^2}2m\l(2m+1\r)},
\end{equation}
and 
\begin{equation}
S_2\simeq {1\over 2}e^{-{2t\over a^2}}+\int_0^\infty dm\l(2m+1\r)e^{-{t\over a^2}\l(2m+1\r)\l(2m+2\r)}.
\end{equation}
By simple changes of variables, we can easily see that
\begin{equation}
S\simeq {1\over 2}\l(1+e^{-{2t\over a^2}}\r)+{1\over 2}\int_0^\infty d\ell \l(2\ell+1\r)e^{-{t\over a^2}\ell\l(\ell+1\r)}-{1\over 2}\int_0^1 d\ell\,\ell e^{-{t\over a^2}\ell\l(\ell+1\r)}.
\end{equation}
Of these terms, the integral from zero to infinity scales as $a^2\over t$, while the other terms scale as order 1 in the $a$ expansion. Hence
\begin{equation}
S\simeq {1\over 2}\int_0^\infty d\ell \l(2\ell+1\r)e^{-{t\over a^2}\ell\l(\ell+1\r)}\simeq {1\over 2}\sum_0^\infty \l(2\ell+1\r)e^{-{t\over a^2}\ell\l(\ell+1\r)},
\end{equation}
where we have approximated the integral over $\ell$ by the corresponding sum. Hence, we have shown that in the large-$a$ approximation, the heat kernel of the scalar Laplacian on $\ss2/\mathbb{Z}_2$ is given by half the heat kernel of the scalar Laplacian in $\ss2$. To complete the proof, we need to show that the error terms $E$ given in \refb{interr} are subleading in the large-$a$ expansion. We shall explicitly demonstrate this for the function which is the summand in $S_1$, i.e.
\begin{equation}
f\l(x\r)=\l(2x+1\r)e^{-{t\over a^2}2x\l(2x+1\r)}.
\end{equation}
Firstly, note that the expression $E$ is finite (in fact, zero) in the limit $t\mapsto 0$. Hence a naive power-series expansion of the terms in $E$ is justified. Now, it is easy to see that $f(x)$ iself is $\mathcal{O}\l(1\r)$ in the small-$t$ (or large-$a$) expansion\footnote{The dimensionless expansion parameter is $t\over a^2$.}. Taking derivatives of $f$ with respect to $x$ will in general pull down more powers of $1\over a^2$, making these terms further suppressed in the large-$a$ approximation. This completes our proof.
\subsubsection{The Euler-Maclaurin Formula and the Heat Kernel on $\sn$}
Now, we return to our original goal of evaluating the heat kernel over the quotient space $\zn$. To do so, we shall use the Euler-Maclaurin Formula, which states that
\begin{equation}\label{eulermac}
\sum_{n=a}^b f\l(n\r)={1\over 2}f\l(a\r)+\int_{a}^b dx\,f(x)+{1\over 2}f\l(b\r)+E,
\end{equation}
where $E$ is given in \refb{interr}. From the analysis of Section \ref{largerad}, it is natural to expect that the heat kernel on $\sn$ should finally be expressible as $1\over N$ times the heat kernel on $\ss2$ with additional terms, which now have to be computed. We will see that, to a given order in $t\over a^2$, only a finite number of terms contribute to $E$, which enables us to explicitly evaluate the heat kernel to that order in $t$. Consider now, the $n$th term in the sum \refb{qsphereN}, $S_n$. Using the Euler-Maclaurin formula, we find that
\begin{equation}
S_n={1\over 2}e^{-t{n(n-1)\over a^2}}+\int_0^\infty dp\l(2p+1\r)e^{-t{(Np+n-1)(Np+n)\over a^2}}-\sum_{k=2}^\infty{B_k\over k!}f_n^{(k-1)}\l(0\r),
\end{equation}
where we have used the fact that the function $f_n(x)=(2x+1)e^{-t{(Nx+n-1)(Nx+n)\over a^2}}$ and its derivatives vanish at $x=\infty$ for arbitrary values of $n$. To solve this further, we substitute $\ell=Np+n-1$ in the integral to find
\begin{equation}
S_n={1\over 2}e^{-t{n(n-1)\over a^2}}+\int_{n-1}^\infty {dl\over N}\l({2\ell+1\over N}+\l(1+{1-2n\over N}\r)\r)e^{-t{\ell(\ell+1)\over a^2}}-\sum_{k=2}^\infty{B_k\over k!}f_n^{(k-1)}\l(0\r).
\end{equation}
This can be further simplified to obtain
\begin{equation}
\begin{split}
S_n=&{1\over 2}e^{-t{n(n-1)\over a^2}}+\int_{0}^\infty {dl\over N^2}\l(2\ell+1\r)e^{-t{\ell(\ell+1)\over a^2}}-\int_{0}^{n-1} {dl\over N^2}\l(2\ell+1\r)e^{-t{\ell(\ell+1)\over a^2}}\\&+\l(1+{1-2n\over N}\r)\int_0^\infty {d\ell\over N}e^{-t{\ell(\ell+1)\over a^2}}-\l(1+{1-2n\over N}\r)\int_0^{n-1}{d\ell\over N}e^{-t{\ell(\ell+1)\over a^2}}-\sum_{k=2}^\infty{B_k\over k!}f_n^{(k-1)}\l(0\r).
\end{split}
\end{equation}
Then the total integrated heat kernel can be obtained by summing $S_n$ over $n$ as follows. Firstly,
\begin{equation}
\sum_{n=1}^N \int_{0}^\infty {dl\over N^2}\l(2\ell+1\r)e^{-t{\ell(\ell+1)\over a^2}}=-{a^2\over Nt},
\end{equation}
and
\begin{equation}
\sum_{n=1}^N \int_{0}^\infty\l(1+{1-2n\over N}\r)\int_0^\infty {d\ell\over N}e^{-t{\ell(\ell+1)\over a^2}}=\l[\int_0^\infty {d\ell\over N}e^{-t{\ell(\ell+1)\over a^2}}\r]\l(0\r)=0.
\end{equation}
The rest of the terms are regular as $t=0$ and we can extract the short time asymptotics by doing a power series expansion of each term. We will retain terms upto $\mathcal{O}(t)$ and discard terms from $\mathcal{O}(t^2)$ onwards. On evaluating the sum $S$ in this approximation, we find
\begin{equation}
S= {1\over N}\int_0^\infty d\ell\l(2\ell+1\r)e^{-t{\ell(\ell+1)\over a^2}}+\frac{N^2+1}{6 N}+\frac{\left(N^4+10 N^2+1\right) t}{180 a^2 N}+\mathcal{O}\l(t^2\r).
\end{equation}
Now, using the Euler-Maclaurin formula to express the above integral in terms of a sum, and again working to linear order in $t\over a^2$, we find that
\begin{equation}\label{heatksn}
S={1\over N}\sum_{\ell=0}^\infty \l(2\ell+1\r)e^{-t{\ell(\ell+1)\over a^2}}+\frac{N^2-1}{6 N}+\frac{\left(N^2+11\right) \left(N^2-1\right) t}{180 a^2 N}+\mathcal{O}\l(t^2\r).
\end{equation}
This is our final result. It expresses the heat kernel for a real scalar field on the quotient space $\sn$ in terms of the heat kernel on the unquotiented sphere, and additional terms, which can be thought of as arising from the fixed points of the $\zn$ orbifold. The coefficient for the $t^0$ term is given for scalars on compact manifolds $\mathcal{M}_\beta$with conical singularities in \cite{Fursaev:1996uz}. Their answer for this term is (see their equations (1.2), (2.7) and (2.8))
\begin{equation}
A_1={1\over 4\pi}\l[{1\over 6}\int_{\mathcal{M}_\beta-\Sigma}R+{1\over 6}\cdot{2\pi\over N}\l(N^2-1\r)(\sharp\Sigma)\r],
\end{equation}
where $\Sigma$ denotes the singular points and $\beta={2\pi\over N}$ is the angle of the cone. For the sphere and its quotients $R={2\over a^2}$, and there are two singular points, the north and south poles of the sphere. Hence the above expression becomes
\begin{equation}
A_1={1\over 4\pi}\cdot{4\pi a^2\over N}\cdot{1\over 3 a^2}+{1\over 4\pi}\cdot{4\pi\over 6}\l(N-{1\over N}\r)={1\over N}\cdot{1\over 3}+{1\over 6}\l({N^2-1\over N}\r)={1\over N}A_1^{S^2}+{1\over 6}\l({N^2-1\over N}\r).
\end{equation}
This matches (term-by-term) with the coefficient of the $t^0$ term which we have obtained above. Here $A_1^{S^2}$ is the coefficient of the $t^0$ term in the heat kernel expansion on $S^2$ without any quotient, and is read off from \refb{A1s2}. 
\subsection{The Analytic Continuation to $\adss/\zn$}\label{scalaranalytic}
The heat kernel for the scalar on the $\zn$ orbifold of $\ss2$ is given by
\begin{equation}
K^s_{\sn}={1\over N}K^s_{\ss2}+\frac{N^2-1}{6 N}+\frac{\left(N^2+11\right) \left(N^2-1\right) t}{180 a^2 N}+\mathcal{O}\l(t^2\r),
\end{equation}
where $K^s_{\ss2}$ denotes the integrated heat kernel on the unquotiented $\ss2$. We will now outline how this answer is to be analytically continued to AdS spaces. We expect that the answers should be related by $a\mapsto ia$, but there are two subtleties. Firstly, the answer on $\adss$ and its $\zn$ orbifold is volume divergent, and hence has to be regulated. The other subtlety is due to the fact that the global properties of the $\zn$ orbifold on $\ss2$ and $\adss$ are slightly different, and this must be accounted for\footnote{Such differences have already been encountered in the cases where the orbifold group acts freely on the sphere and hyperboloid\cite{David:2009xg,Gopakumar:2011qs}. The analytic continuation acress the quotient spaces should account for these differences, but once that is done, completely consistent answers are obtained. See for example \cite{Gaberdiel:2010ar,Gupta:2012he,Lal:2012ax}.}. We first show how the differences in the global properties of the quotient group action may be 
accounted for. The discussion in \cite{Fursaev:1996uz} will be useful for this purpose. Consider the integrated heat kernel over a manifold $\mathcal{M}_\beta$ with conical singularities located at points $p_i$. Further, let $\epsilon_i$ be (arbitrarily) small disconnected regions enclosing points $p_i$. Then the integrated heat kernel over $\mathcal{M}_\beta$ may be decomposed into integrals over the small neighbourhoods of the singular points and an integral over the rest of the manifold, which is smooth.
\begin{equation}
\int_{\mathcal{M}_\beta}K\l(x,x,t\r)=\int_{\mathcal{M}_\beta-\lbrace\bigcup\epsilon_i\rbrace}K\l(x,x,t\r)+\sum_i\int_{\epsilon_i} K\l(x,x,t\r).
\end{equation}
Consider now our answer \eqref{heatksn} for the quotiented sphere $S^2\backslash N$. The first term corresponds to the integral over the remaining smooth manifold, and the rest of the terms arise from integrating about the conical singularities located at the north and south poles of the sphere. We argue this as follows: The integrated heat kernel on the smooth manifold $\mathcal{M}_\beta-\lbrace\bigcup\epsilon_i\rbrace$ admits an expansion in powers of $t$ where the coefficients are expressible in terms of volume integrals of local general-coordinate invariant quantitites \cite{Vassilevich:2003xt}. As $S^2$ is homogeneous, these invariants are independent of the location on $S^2$ (or its quotients), and these integrals on such manifolds are just the volume of the manifold times a constant. On $\sn$ (once the singular points, the north and south poles, have been removed), therefore, the answer is ${1\over N}$ times the answer on $S^2$. As there are two conical singularities on S$^2$, each cone contributes
\begin{equation}\label{conecontrib}
{1\over 2}\l[\frac{N^2-1}{6 N}+\frac{\left(N^2+11\right) \left(N^2-1\right) t}{180 a^2 N}+\mathcal{O}\l(t^2\r)\r].
\end{equation}
Now the AdS$_2$ under this quotient has only one fixed point. It should again be possible to carry out the above decomposition of the integration domains into the smooth part and a small neighbourhood containing the singularity. 
\begin{equation}
\int_{AdS_2\backslash N}K\l(x,x,t\r)=\int_{AdS_2\backslash N-\epsilon}K\l(x,x,t\r)+\int_{\epsilon} K\l(x,x,t\r),
\end{equation}
where $\epsilon$ contains the origin, where the conical singularity is located. By the above considerations, we analytically continue the conical contribution to the heat kernel \eqref{heatksn} on $\sn$ as
\begin{equation}
K_{\adsn}^{\text{conical}}={1\over 2}\l[\frac{N^2-1}{6 N}-\frac{\left(N^2+11\right) \left(N^2-1\right) t}{180 a^2 N}\r]+\mathcal{O}\l(t^2\r),
\end{equation}
where we have analytically continued $a\mapsto ia$ as well. The analytic continuation of the first term in \refb{heatksn} is well understood \cite{Camporesi:1994ga}. Since both $\ss2$ and $\adss$ are homogeneous spaces, we have
\begin{equation}
K^s_{\ss2}=\text{Vol.}_{\ss2} K_1^s\l(a\r),\quad K^s_{\adss}=\text{Vol.}_{\adss} K_2^s\l(a\r),
\end{equation}
where $K_1^s$ and $K_2^s$ are the coincident scalar heat kernels over $\ss2$ and $\adss$ respectively. These are related by analytic continuation\footnote{We refer the reader to \cite{Banerjee:2010qc} for explicit expressions. We also emphasize that there the coincident heat kernel on $\adss$ is computed independently of this analytic continuation.}. In particular,
\begin{equation}\label{logsphere1}
K_1^s\l(a\r)={1\over 12\pi a^2}+{1\over 4\pi t}+{t\over 60\pi a^4}+\mathcal{O}(t^2),
\end{equation}
and
\begin{equation}\label{logads1}
K_2^s\l(a\r)=-{1\over 12\pi a^2}+{1\over 4\pi t}+{t\over 60\pi a^4}+\mathcal{O}(t^2).
\end{equation}
A derivation of \eqref{logsphere1}, again employing the Euler-Maclaurin formula \eqref{eulermac}, is presented in Appendix \ref{A}. The final step is to regulate the volume divergence of $\adss$. This is again well understood by now. However, since we will use this procedure to regulate all the divergences we encounter in our analysis, we will review this briefly here. We state the main steps, referring the reader to \cite{Sen:2008vm,Sen:2009vz,Banerjee:2010qc,Bhattacharyya:2012ye} for details. The $\adss$ volume integral is given by
\begin{equation}
V=\int_0^\infty\int_0^{2\pi} d\eta d\phi \l(a^2\sinh\eta\r),
\end{equation}
which clearly diverges. To regulate this divergence, we put a cutoff on $\eta$ at a large value $\eta_0$, which gives the regulated volume of $\adss$ to be
\begin{equation}
V_{\text{reg}}=2\pi a^2 \l(\cosh\eta_0-1\r)=2\pi a^2 \l({1\over 2}e^{\eta_0}-1+\mathcal{O}\l(e^{-\eta_0}\r)\r).
\end{equation}
Now, following the discussion in \cite{Bhattacharyya:2012ye}, since the divergent term $e^{\eta_0}$ may be expressed in terms of the radius of curvature of the boundary sphere, it can be cancelled off by boundary counterterms. Hence, the regularised volume of $\adss$ is given by
\begin{equation}\label{regvol}
V=-2\pi a^2.
\end{equation}
Hence, the integrated scalar heat kernel on $\adsn$ is given by
\begin{equation}\label{scalarads2n}
K^s_{\adsn}={1\over N}\l(-{a^2\over 2 t}+{1\over 6}-{t\over 30 a^2}\r)+{1\over 2}\l[\frac{N^2-1}{6 N}-\frac{\left(N^2+11\right) \left(N^2-1\right) t}{180 a^2 N}\r]+\mathcal{O}\l(t^2\r),
\end{equation}
where the volume divergence has been regulated as above.
\subsubsection{A Check of The Analytic Continuation}
In the previous section, we have shown how the heat kernel for the scalar on $\sn$ may be analytically continued to obtain the heat kernel on $\adsn$. In this section, we will verify this explicitly against expressions obtained in the existing literature. In particular, the scalar heat kernel was computed on the $\adss$ cone in \cite{Mann} using the Sommerfeld formula \cite{sommer1,sommer2,sommer3}. It was found that if the angular coordinate of $\adss$ is quotiented by $\theta\mapsto\theta+2\pi\alpha$, where $\alpha\in\mathbb{R}_+$, then the traced heat kernel on the quotient space (called $H_2^\alpha$ in \cite{Mann}) is given by
\begin{equation}\label{sommerkernel}
\tr K_{H_2^\alpha}=\alpha\tr K_{H_2}+\alpha{e^{-\bar{s}\over 4}\over\l(4\pi s\r)^{1\over 2}} \int_0^\infty dy\cosh y f\l(y,\alpha\r)e^{-y^2\over\bar{s}},
\end{equation}
where $\bar{s}={t\over a^2}$ and the function $f(y,\alpha)$ was determined to be
\begin{equation}
f\l(y,\alpha\r)= {1\over\sinh^2 y}\l(1-{2y\over\sinh 2y}\r)\l(1-\alpha\r) +\mathcal{O}\l(\l(1-\alpha\r)^2\r).
\end{equation}
We refer the reader to equations (20) to (23) of \cite{Mann} for more details regarding these expressions. Now in the $\alpha\simeq 1$ expansion, the expression \eqref{sommerkernel} becomes
\begin{equation}\begin{split}
\tr K_{H_2^\alpha}&=\alpha\tr K_{H_2} +{\alpha\over 2}{e^{-{\bar{s}\over 4}}}\l({1\over 3}-{\bar{s}\over 20} +\mathcal{O}\l(\bar{s}^2\r)\r)\l(1-\alpha\r)\\&\simeq \alpha\tr K_{H_2} +{\alpha\l(1-\alpha\r)\over 2}\l({1\over 3}-{2\bar{s}\over 15}+\mathcal{O}\l(\bar{s}^2\r)\r).
\end{split}
\end{equation}
We now identify the parameter $\alpha$ with our ${1\over N}$, and find that
\begin{equation}
\tr K_{H_2^N}\simeq {1\over N}\tr K_{H_2} +\l(N-1\r)\l({1\over 6}-{1\over 15}{t\over a^2}+\ldots\r),
\end{equation}
where we have retained the leading term in the $N\approx 1$ expansion. This is precisely the expression obtained on expanding \eqref{scalarads2n} formally around $N = 1$.
\subsection{The Hodge Laplacian for Vector Fields}\label{hodgelap}
Let us consider, following \cite{Banerjee:2010qc}, the path integral for a U(1) gauge field in the Euclidean signature on a background manifold $\mathcal{M}$, equipped with a metric $g$.
\begin{equation}
S_A=\int d^d{x}\sqrt{g} F_{\mu\nu}F^{\mu\nu},
\end{equation}
where $F_{\mu\nu}$ is the U(1) field strength $\partial_\mu A_\nu-\partial_\nu A_\mu.$ To evaluate the path integral, we will fix gauge by adding the term
\begin{equation}
S_{gf}=-{1\over 2}\int d^dx\sqrt{g} \l(D_\mu A^\mu\r)^2,
\end{equation}
so that the total action becomes
\begin{equation}
S=S_A+S_{gf}=-{1\over 2}\int d^dx\sqrt{g}A_\mu\l(\Delta A\r)^\mu,
\end{equation}
where 
\begin{equation}
\l(\Delta A\r)_\mu= -\square A_\mu +R_{\mu\nu} A^\nu,\quad \square A_\mu=g^{\rho\sigma}D_\rho D\sigma A_\mu
\end{equation}
is the Hodge Laplacian, expressed as 
\begin{equation}
\Delta\equiv d\delta+\delta d,
\end{equation}
where $\delta=-*d*$, and $*$ is the Hodge dual defined with respect to the background metric $g$. This gauge fixing requires us to introduce two anticommuting scalar ghost fields whose kinetic operator is the usual Laplacian. Then the one-loop contribution of the gauge field is given by
\begin{equation}
{1\over 2}\int_\epsilon^\infty {ds\over s}\int d^dx\sqrt{g}\l[K^v\l(x,x,s\r)-2K^s\l(x,x,s\r)\r].
\end{equation}
The first term is the heat kernel of $\Delta$ over vector fields, while the second term represents the contribution due to the ghost fields. Now, in the rest of this section, we will consider the case where $\mathcal{M}$ is either $\ss2$ or $\adss$, or their $\zn$ orbifolds. We will use the fact that, upto global issues, the vector field on $\adss$ and $\ss2$ may be written as
\begin{equation}\label{gaugefielddecomposition}
A_\mu=\nabla_\mu\phi_1+\epsilon_{\mu\nu}\nabla^\nu\phi_2,
\end{equation}
where $\phi_1$ and $\phi_2$ are scalars on the corresponding manifold. For the spaces $\ss2$ and $\adss$, one can show that if \footnote{Though $\nabla$s don't commute over arbitrary tensors, they do commute over scalars, and hence $\epsilon\nabla\phi$ is a transverse vector mode. Further, as $\epsilon_{\mu\nu}$ is covariantly constant, we will concentrate on just $\nabla_{\mu}\phi_1$, and our discussion goes through for transverse modes as well.}
\begin{equation}
\Delta\nabla_\mu\phi_1=\Lambda\nabla_\mu\phi_1,
\end{equation}
where $\Delta=d\delta+\delta d$ is the Hodge Laplacian over 1-forms, then we must have
\begin{equation}\label{vecscalrel2}
\square\phi_1=\l(\Lambda\mp{1\over a^2}\r)\phi_1.
\end{equation}
Hence, eigenmodes of the vector Laplacian are in one-to-one correspondence with scalar eigenfunctions. Hence, the spectrum and degeneracies for the vector Laplacian can be obtained from those of the scalar Laplacian, and one would naively conclude
\begin{equation}
K^v= K^{v_T}+K^{v_L}= K^{s}+K^{s}=2 K^s,
\end{equation}
for both $\ss2$ and $\adss$ and $v_T$ and $v_L$ are longitudinal and transverse modes of the gauge field. However, there are corrections to this expression in both the $\ss2$ and $\adss$ cases. In the $\ss2$ case, there is a correction due to the fact that the $\ell=0$ mode of the scalar is a constant over the two-sphere and does not give rise to a non-trivial gauge field. We account for this by explicitly starting the sum over scalar eigenvalues from $\ell=1$ when evaluating the heat kernel. So the heat kernel on $\ss2$ for the Hodge Laplacian over vectors is given by
\begin{equation}
K^v\l(t\r)=2\sum_{\ell=1}^\infty \l(2\ell+1\r) e^{-t\l({\ell(\ell+1)\over a^2}\r)}=2\l(K^s_{\ss2}-1\r).
\end{equation}
The case of $\adss$ and its quotients is more subtle. It has to do with the presence of harmonic 1-forms in the spectrum of the Hodge Laplacian which lead to zero modes in the full $\adss\otimes\ss2$ geometry and its $\zn$ orbifold. These modes cannot be reached by the analytic continuation we are working with. We shall come to these modes in the next subsection, but for the moment we note that for transverse and longitudinal modes of the vector field on $\adss$ there is no subtlety, and the heat kernel for these fields equals just the heat kernel of the scalar Laplacian. 
\begin{equation}
K^{v_T}_{\adss}=K^{v_L}_{\adss}=K^s_{\adss}.
\end{equation}
In particular, there is no need to subtract out the constant mode of the scalar field as it is not a normalisable mode on $\adss$ \cite{Banerjee:2010qc}. The discussion of the corresponding quotient spaces is entirely analogous. We finally find that
\begin{equation}
K^{v_T}_{\sn}=K^{v_L}_{\sn}=K^{s}_{\sn}-1, \quad K^{v}_{\sn}=2\l(K^s_{\sn}-1\r),
\end{equation}
and
\begin{equation}
K^{v_T}_{\adsn}=K^{v_L}_{\adsn}=K^{s}_{\adsn}.
\end{equation}
For the vector field on $\sn$, we can compute the $\mathcal{O}\l(t^0\r)$ term in the heat kernel expansion. We find that the conical singularities contribute
\begin{equation}
2{N^2-1\over 6 N}+2\l({1\over N}-1\r)
\end{equation}
to the $t^0$ term, which precisely matches with the expression obtained by the methods of \cite{Fursaev:1996uz}. We close this section by a brief discussion of the analytic continuation of the vector heat kernel from $\ss2$ and $\sn$ to $\adss$ and $\adsn$. Firstly, vector fields on $\adss$ and $\adsn$ have longitudinal and transverse modes as above, in addition, there is a series of harmonic modes as well \cite{Camporesi:1994ga}. These arise from field configurations
\begin{equation}
A=d\Phi,
\end{equation}
which are square-integrable, though $\Phi$ itself is not square-integrable. These configurations are not captured by analytic continuation from the sphere, but also contribute to the gauge field path integral. We shall treat these configurations in detail in Section \ref{zeromodes}. For the moment, we concentrate on the longitudinal and transverse modes of the gauge field. In this case, the natural analytic continuation is to analytically continue from $K^{v_T}_{\sn}$ and $K^{v_L}_{\sn}$ after adding the $\ell=0$ mode back to the heat kernel. This of course is equivalent to evaluating the Laplacian over the full set of modes of the scalar fields $\phi_1$ and $\phi_2$ in \eqref{gaugefielddecomposition}, and we indeed find after this analytic continuation that 
\begin{equation}
K^{v_{T,L}}_{\adsn}=K^{s}_{\adsn}.
\end{equation}
This is the procedure we shall follow for the analytic continuation when we take into account the graviphoton background in Section \ref{graviphoton}.
\section{The One-loop determinants on $\ads2s2n$}\label{lapkernelads2s2n}
In this section we will compute the one-loop determinants of Laplacian over the scalar, vector and Dirac spinor field in the $\ads2s2n$ geometry. We will do this by computing the corresponding determinant over $\s2s2n$ geometry and analytically continuing the answer to the AdS case. This will serve as a useful warm-up for the final computation of Section \ref{graviphoton}, where the effects of the graviphoton coupling to the $\cn=4$ vector multiplet fields is accounted for. Additionally, we will find a particularly suggestive form of the answer closely related to those obtained previously for freely acting quotients of spheres and hyperboloids in \cite{David:2009xg,Gopakumar:2011qs}. We begin with defining the geometry and the orbifold projection that will be imposed. In global coordinates, $\adss\otimes\ss2$ may be written as 
\begin{equation}\label{adsmetric}
ds^2=a^2\left(d\eta^2+\sinh^2\eta d\theta^2\right)+a^2\left(d\rho^2+\sin^2\rho d\phi^2\right),
\end{equation}
where $\left(\eta,\theta\right)$ are coordinates on $AdS_2$ and $\left(\rho,\phi\right)$ are coordinates on $S^2$. We will consider, in what follows, the $\zn$ orbifold of this geometry
\begin{equation}\label{orbact}
\l(\theta,\phi\r) \mapsto \l(\theta+{2\pi \over N},\phi-{2\pi \over N}\r).
\end{equation}
In particular, we will compute the heat kernel of the Laplacian for various spin-fields over the modes in $\adss\otimes\ss2$ that are left invariant under the orbifold action. We note here that this geometry may be reached from the space $\s2s2n$ with metric 
\begin{equation}\label{s2metric}
ds^2=a_1^2\l(d\chi^2+\sin^2\chi d\theta^2\r)+a_2^2\l(d\rho^2+\sin^2\rho d\phi^2\r),
\end{equation}
via the analytic continuation $a_1\mapsto ia$, $a_2\mapsto a$, $\chi\mapsto i\eta$. The orbifold action on $\ss2\otimes\ss2$ is still given by \refb{orbact}. We now turn to the evaluation of the integrated heat kernel over these two geometries. Our strategy will be to write down the full sprectrum of the Laplacian over the unquotiented geometries $\ss2\otimes\ss2$ and $\adss\otimes\ss2$, and impose an orbifold projection which maps to the subset of modes invariant under \eqref{orbact}. We shall then explicitly evaluate the heat kernel over this subset of modes. We mostly obtain explicit answers for the compact space $\s2s2n$, and indicate how the answer may be analytically continued to $\ads2s2n$.
\subsection{The Scalar Field}\label{scalarprojectionmethod}
The eigenfunctions of the Laplacian acting over scalar field in $\adss\otimes\ss2$ are obtained by tensoring the scalar eigenfunctions on $\ss2$ with the scalar eigenfunctions on $\adss$. That is, the spectrum of scalar eigenfunctions on $\adss\otimes\ss2$ is given by
\begin{equation}\label{scalareigenf}
\Phi_{\lambda,m,\ell,n}\l(\eta,\theta,\rho,\phi\r)=f_{\lambda,m}\l(\eta,\theta\r)Y_{\ell,n}\l(\rho,\phi\r),
\end{equation}
where the $\Psi_{\lambda,m}$ are the eigenfunctions of the scalar Laplacian on $\adss$, given--upto normalisation--by \cite{Camporesi:1994ga}
\begin{equation}\begin{split}
f_{\lambda,m}\l(\eta,\theta\r)=\sinh^{\vert m\vert}\eta{_2F_1}\l(i\lambda+\vert m\vert+{1\over 2},-i\lambda\r.&+\l.\vert m\vert +{1\over 2},\vert m\vert +1,-\sinh^2{\eta\over 2}\r)e^{im\theta},\\& 0<\lambda<\infty,\quad m\in\mathbb{Z},
\end{split}\end{equation}
and the $Y_{\ell,n}$s are the usual spherical harmonics on $\ss2$. The eigenfunction \refb{scalareigenf} belongs to the eigenvalue 
\begin{equation}\label{scalareigenv}
E_{\lambda,\ell}={1\over a^2}\l(\lambda^2+{1\over 4}+\ell\l(\ell+1\r)\r).
\end{equation}
The eigenfunctions of the scalar Laplacian on $\ss2\otimes\ss2$ are given by 
\begin{equation}\label{s2scalef}
\Psi_{\tilde{\ell},m,\ell,n}\l(\chi,\theta,a_1,\rho,\phi,a_2\r)=Y_{\tilde{\ell},m}\l(\chi,\theta,a_1\r)Y_{\ell,n}\l(\rho,\phi,a_2\r),
\end{equation}
which belong to the eigenvalue 
\begin{equation}
E_{\tilde{\ell},\ell}={1\over a_1^2}\tilde{\ell}\l(\tilde{\ell}+1\r)+{1\over a_2^2}\ell\l(\ell+1\r).
\end{equation}
Now, we will use the fact that the spectrum of the scalar Laplacian on $\adss\otimes\ss2$ is related to the spectrum of the scalar Laplacian on $\ss2\otimes\ss2$ by the analytic continuation \cite{Camporesi:1994ga}
\begin{equation}
\tilde{\ell}\mapsto i\lambda-{1\over 2},\, a_1\mapsto ia,\, \chi\mapsto i\eta,\, a_2\mapsto a.
\end{equation}
In particular, we will impose the projection \refb{orbact} on the modes on the compact space $\ss2\otimes\ss2$ and compute the integrated heat kernel by enumerating the eigenvalues and their degeneracies. Then, we will obtain the answer over $\ads2s2n$ by analytic continuation. We now begin with imposing the orbifold action \refb{orbact} on modes \refb{s2scalef}. Modes invariant under the orbifold action obey the quantisation condition
\begin{equation}
\label{cons1}
 m-n =N p\qquad, p\in \mathbb{Z}_+.
\end{equation}
Firstly, note that the unconstrained integrated heat kernel on $\ss2\otimes\ss2$ may be expressed as
\begin{equation}
K^s= K^s_{\ss2}\cdot K^s_{\ss2}= \sum_{\ell=0}^\infty\sum_{\tilde{\ell}=0}^\infty \sum_{m=-\tilde{\ell}}^{\tilde{\ell}} \sum_{n=-\ell}^{\ell} e^{-{t\over a_1^2}\tilde{\ell}\l(\tilde{\ell}+1\r)}e^{-{t\over a_2^2}\ell\l(\ell+1\r)}.
\end{equation}
From this, we can extract the heat kernel of the states invariant under the projection \refb{orbact} by inserting a Kronecker delta over $m$ and $n$ which is nonvanishing when $m$ and $n$ obey \refb{cons1}. We choose the following representaion for this delta function
\begin{equation}
\delta_{m,n,N}={1\over N}\sum_{s=0}^{N-1} e^{2\pi i (m-n) {s\over N}},\quad m,\, n,\, s \in \mathbb{Z}.
\end{equation}
Then the scalar heat kernel over the orbifold geometry $\s2s2n$ is given by
\begin{equation}\label{scalarkernel}
K^s = {1\over N} \sum_{s=0}^{N-1} \left[\left(\sum_{\tilde{\ell}=0}^{\infty} \frac{\sin\left[\frac{\pi (2\tilde{\ell}+1 ) s}{N}\right]}{\sin\left[\frac{\pi  s}{N}\right]} e^{-{t\over {a_1^2}}\tilde{\ell}(\tilde{\ell}+1)}\right)\left(\sum_{\ell=0}^{\infty} \frac{\sin\left[\frac{\pi (2\ell+1) s}{N}\right]}{\sin\left[\frac{\pi s}{N}\right]} e^{-{t\over {a_2^2}}\ell(\ell+1)}\right)\right],
\end{equation}
which has been obtained by carrying out the sums over $m$ and $n$. We recognize the functions 
\begin{equation}
\chi_{\ell}\l(s\r)= \frac{\sin\left[\frac{\pi (2\ell+1) s}{N}\right]}{\sin\left[\frac{\pi s}{N}\right]}
\end{equation}
as the characters in the spin-$\ell$ representation of $SU(2)$. The above answer may be expressed in the following form, which suggests an analytic continuation to AdS.
\begin{equation}\label{scalars2s2n}
K^s= {1\over N} K^s_{\ss2\otimes\ss2}+{1\over N} \sum_{s=1}^{N-1} \left[\left(\sum_{\tilde{\ell}=0}^{\infty} \chi_{\tilde{\ell}}\l(s\r) e^{-{t\over {a_1^2}}\tilde{\ell}(\tilde{\ell}+1)}\right)\left(\sum_{\ell=0}^{\infty}\chi_{\ell}\l(s\r) e^{-{t\over {a_2^2}}\ell(\ell+1)}\right)\right].
\end{equation}
Here $K^s_{\ss2\otimes\ss2}$ is the integrated heat kernel of the scalar over the unquotiented space $\ss2\otimes\ss2$. We will now perform an analytic continuation of this answer to the quotient space $\ads2s2n$. Firstly, we note that $K^s_{\ss2\otimes\ss2}$ may be expressed as 
\begin{equation}\label{scalars2s2}
K^s_{\ss2\otimes\ss2}\l(a_1,a_2\r)=K^s_{\ss2}\l(a_2\r)\sum_{\tilde{\ell}=0}^\infty{\l(2\tilde{\ell}+1\r)} e^{-{t\over a_1^2}\tilde{\ell}\l(\tilde{\ell}+1\r)}.
\end{equation}
The first term is just the heat kernel for the scalar on the $\ss2$ that acts as a spectator, its analytic continuation is trivial, just the radius $a_2$ has to be replaced by $a$. The second term is the heat kernel on the $\ss2$ which is being analytically continued to $\adss$. The rules for this analytic continuation have been very precisely specified in Section \ref{scalaranalytic}. Finally, the second term is \refb{scalars2s2} may be regarded as arising due to the presence of fixed points in our quotient space. This is analytically continued using our usual prescription of $a_1\mapsto ia$, and multiplying an overall factor of ${1\over 2}$ to account for the global feature that $\adss$ has only one fixed point under the action of this orbifold, while the sphere, from which we're doing this analytic continuation has two. For the specific example of the $\mathbb{Z}_2$ orbifold, this leads to the final expression 
\begin{equation}
K^s_{{\adss\otimes\ss2}/{\mathbb Z}_2}=\frac{1}{2}K^s_{\adss}K^s_{\ss2}+\frac{1}{4}\left[\frac{1}{4}+{\mathcal O}(t^2)\right].
\end{equation}
\subsection{The Vector Field}
We now turn to the evaluation of the heat kernel of the Hodge Laplacian over the vector field on $\adss\otimes\ss2$ and its $\zn$ orbifold. This may be decomposed into
\begin{equation}
K^v_{\ads2s2n}=K^v_{\l(\adss\r.}K^s_{\l.\ss2\r)/\zn}+K^s_{\l(\adss\r.}K^v_{\l.\ss2\r)/\zn}
\end{equation}
To begin with, we consider the heat kernel evaluated purely over longitudinal and transverse modes of the gauge field on $\adsn$. We will account for the presence of harmonic modes subsequently. For the non-zero modes on $\adsn$, the contribution to the heat kernel on $\ads2s2n$ may be arrived at by analytic continuation from the vector field heat kernel on $\s2s2n$. We will first consider the case of $N=1$, i.e. the unquotiented spaces and specify the analytic continuation. The heat kernel on $\ss2\otimes\ss2$ is given by
\begin{equation}
K^{v}_{\ss2\otimes\ss2}\l(a_1,a_2\r)=K^{v}_{\ss2}\l(a_1\r)K^{s}_{\ss2}\l(a_2\r)+K^{s}_{\ss2}\l(a_1\r) K^{v}_{\ss2}\l(a_2\r),
\end{equation}
which may be written out as
\begin{equation}\begin{split}
K^{v}_{\ss2\otimes\ss2}\l(a_1,a_2\r)=\l[2\sum_{\tilde{\ell}=1}^\infty\r. &\l. \l(2\tilde{\ell}+1\r) e^{-t\l({\tilde{\ell}(\tilde{\ell}+1)\over a_1^2}\r)}\r]\l[\sum_{\ell=0}^\infty \l(2\ell+1\r) e^{-t\l({\ell(\ell+1)\over a_2^2}\r)}\r]\\&+\l[\sum_{\tilde{\ell}=0}^\infty \l(2\tilde{\ell}+1\r) e^{-t\l({\tilde{\ell}(\tilde{\ell}+1)\over a_1^2}\r)}\r]\l[2\sum_{\ell=1}^\infty \l(2\ell+1\r) e^{-t\l({\ell(\ell+1)\over a_2^2}\r)}\r].
\end{split}\end{equation}
This may be re-expressed as 
\begin{equation}
K^{v}_{\ss2\otimes\ss2}\l(a_1,a_2\r)=2\l(K^{s}_{\ss2}\l(a_1\r)-1\r)K^{s}_{\ss2}\l(a_2\r) +2K^{s}_{\ss2}\l(a_1\r)\l(K^{s}_{\ss2}\l(a_2\r)-1\r).
\end{equation}
Now, we will analytically continue this to the $\adss$ case. This involves the following replacements. Firstly, the heat kernel of the scalar on the $\ss2$ of radius $a_1$ gets continued to the heat kerne of the scalar on $\adss$ of radius $a$ as above. Secondly, there is no need to subtract out the constant mode of the scalar in the first term. We thus obtain
\begin{equation}
K^{v}_{\adss\otimes\ss2}\l(a\r)=2 K^{s}_{\adss}\l(a\r)K^{s}_{\ss2}\l(a\r) +2K^{s}_{\adss}\l(a\r)\l(K^{s}_{\ss2}\l(a\r)-1\r).
\end{equation}
The above discussion goes through on the $\zn$ quotients as well, where we have to be careful to project onto the appropriate subset of modes invariant under the orbifold, but no other change is required. In this case, it is apparent that the final answer on $\s2s2n$ is given by
\begin{equation}
K^{v}_{\s2s2n}=4K^{s}_{\s2s2n}\l(a_1,a_2\r)-2K^{2}_{\sn}\l(a_1\r)-2K_{\sn}\l(a_2\r).
\end{equation}
That the last two terms are the heat kernel evaluated on the modes invariant under the orbifold projection on a single $\ss2$ should be apparent from the fact that the constant mode on one sphere corresponds to modes $(m,n)=(0,n)$, in which case the modes invariant under \refb{orbact} are $\l(m,n\r)=\l(0,Np\r)$, where $p\in\mathbb{Z}$. These are just modes invariant under the $\zn$ orbifold on the other $\ss2$. Now the analytic continuation to $\ads2s2n$ may be obtained from continuing the scalar heat kernels, and not subtracting out the constant mode on $\ss2\l(a_1\r)$. We then obtain
\begin{equation}
K^{v}_{\ads2s2n}=4K^{s}_{\ads2s2n}\l(a,a\r)-2K^{2}_{\adss}\l(a\r).
\end{equation}
We remind the reader that this expression is the heat kernel for the transverse and longitudinal modes of the vector field. We have not included the harmonic modes in this analysis. Neither have we subtracted out the ghost determinant which arises out of gauge fixing. 
\subsubsection{Zero Modes of the Vector Field in $\adss$}\label{zeromodes}
Now we take into account the presence of modes on $\adss$ which are not related by analytic continuation to $\ss2$. These are
\begin{equation}\label{adszero}
f_a^m=\nabla_a\phi^m,\quad \phi^m={\sqrt{1\over 2\pi\vert m\vert}}\left[\sinh\eta\over 1+\cosh\eta\right]^{\vert m\vert}e^{i m\theta},\, m\in \mathbb{Z}-\lbrace0\rbrace.
\end{equation}
These contribute in two ways. Firstly, they tensor with non-zero modes of the scalar on $\ss2$ to produce non-zero modes of the vector field in $\adss\otimes\ss2$ and its $\zn$ orbifold. Secondly, they tensor with the zero mode of the scalar on $\ss2$ to produce zero modes in $\adss\otimes\ss2$ and its $\zn$ orbifold. We will first calculate the heat kernel contribution of the modes \refb{adszero} to the non-zero modes
\begin{equation}\begin{split}
K^{b'}&= \sum_{m\in\mathbb{Z}-\lbrace 0\rbrace}\sum_{\ell=1}^\infty \sum_{n=-\ell}^{\ell}\int^{\eta_0}\sqrt{g}d\eta d\theta g^{ab} f_a^{* m}f_b^{m} \delta_{m-n,Np} e^{-{t\over a^2}\ell(\ell+1)}\\& = {1\over N}\sum_{s=0}^{N-1} \l(\sum_{m\in\mathbb{Z}-\lbrace 0\rbrace} \tanh\l({\eta_0\over 2}\r)^{2\vert m\vert}e^{im {2\pi s\over N}}\r)\sum_{\ell=1}^\infty \chi_{\ell}\l({\pi s \over N}\r)e^{-{t\over a^2}\ell(\ell+1)}.
\end{split}\end{equation}
We will now explicitly evaluate this expression. The bracketed term sums to
\begin{equation}
T(s)={1\over 1-\tanh^2\l({\eta_0\over 2}\r)e^{{2\pi i s\over N}}}-1+{1\over 1-\tanh^2\l({\eta_0\over 2}\r)e^{-{2\pi i s\over N}}}-1.
\end{equation}
Further, we can show that
\begin{equation}
T(0)\simeq {1\over 2}e^{\eta_0}-1+\mathcal{O}\l(e^{-\eta_0}\r),\quad T\l(m\r)\simeq-1+\mathcal{O}\l(e^{-\eta_0}\r), 1\leq m\leq N-1.
\end{equation}
Then, in accordance with our usual rule for regularising these divergences, we retain the order-1 term only to find that
\begin{equation}\label{znonz}
K^{b'}=-1\l[{1\over N}\sum_{s=0}^{N-1}\sum_{\ell=1}^\infty\chi_{\ell}\l({\pi s \over N}\r)e^{-{t\over a^2}\ell(\ell+1)}\r]=-K^s_{\sn}+1.
\end{equation}
Now, to compute the contribution of the zero modes to the vector field, it is sufficient to compute the number of zero modes. On non-compact spaces like $AdS_2/{\mathbb Z}_N$ and $\ads2s2n$, this is typically a divergent quantity which has to be regulated. We now describe how this is done. We take our definition of the number of zero modes to be
\begin{equation}\label{zerodef}
n_0 =\sum_{p\in \mathbb{Z}-\lbrace 0\rbrace}\int d\theta d\eta\sqrt{g} g^{mn} f^{\ell *}_m f_n^{\ell},
\end{equation}
where $\ell=Np$ for $p\in\mathbb{Z}$.
Our strategy will be to compute the above integral by cutting off the AdS radial coordinate $\eta$ at some large value $\eta_0$ to compute the above sum (which is completely convergent in that regularisation). Then we will do the large $\eta_0$ expansion to pick out the order 1 term which will be our definition of the number of zero modes. This is completely equivalent to the way we compute zero modes on the homogeneous spaces like $\adss$ where the integral (essentially the volume of the AdS space) is regularised in this manner. 
\begin{eqnarray}
n_0 &=&\sum_{p\in \mathbb{Z}-\lbrace 0\rbrace}\int_{AdS_2}^{\eta_0} d\theta d\eta\sqrt{g} g^{mn} f^{Np *}_m f_n^{Np}=\sum_{p\in \mathbb{Z}-\lbrace 0\rbrace}(\tanh\frac{\eta_0}{2})^{2N|p|}\\=&& 2\frac{(\tanh\frac{\eta_0}{2})^{2N}}{1-(\tanh\frac{\eta_0}{2})^{2N}}\simeq {1\over 2N}e^{\eta_0}-1+\mathcal{O}\l(e^{-\eta_0}\r),
\end{eqnarray}
in the large $\eta_0$ limit. We drop the factor diverging with the AdS radial coordinate, and keep the order 1 term as the number of zero modes. Hence the number of zero modes in the quotient space $\adsn$ is given by
\begin{equation}\label{nzero}
n^0_{\adsn}=-1.
\end{equation}
In Appendix \ref{zetafunc} we outline how a zeta function regularization of \eqref{zerodef} gives the same answer as obtained here. Finally, zero modes on the space $\ads2s2n$ are obtained by tensoring the harmonic modes \ref{adszero} on $\adss$ with the constant mode of the scalar on $S^2$. It is straightforward to check from the orbifold projection \ref{orbact} that the only field configurations that can contribute are harmonic modes of vector fields on $\adss$ with moding $m=Np$, $p\in\mathbb{Z}$. Hence, we obtain
\begin{equation}
n^0_{\ads2s2n}=n^0_{\adsn}=-1.
\end{equation}
\subsection{The Dirac Spinor}
We finally examine how the above discussion extends to the case of the Dirac spinor on $S^2\times S^2$, $AdS_2\times S^2$ and their $\zn$ quotients. We will again compute the answer explicitly on the compact space $\s2s2n$ and analytically continue the answer to $\ads2s2n$. We first recollect that the eigenstates of $\slashed{D}_{S^2}$ are given by
\begin{equation}
\chi_{l,m}^{\pm} = {1\over \sqrt{4\pi a^2}}\,
{\sqrt{(l-m)!(l+m+1)!}\over l!}\,
e^{ i\left(m+{1\over 2}\right)\phi} 
\left(\begin{array}{cc} i\, \sin^{m+1}{\psi\over 2}\cos^m {\psi\over 2}
P^{\left(m+1, m\right)}_{l-m}(\cos\psi) \\ \pm \sin^{m}{\psi\over 2}\cos^{m+1} {\psi\over 2}
P^{\left(m, m+1\right)}_{l-m}(\cos\psi) \end{array}\right),
\end{equation}
and
\begin{equation}
\eta_{l,m}^{\pm} = {1\over \sqrt{4\pi a^2}}\,
{\sqrt{(l-m)!(l+m+1)!}\over l!}\,
e^{ -i\left(m+{1\over 2}\right)\phi} 
\left(\begin{array}{cc} i\, \sin^{m}{\psi\over 2}\cos^{m+1} {\psi\over 2}
P^{\left(m+1, m\right)}_{l-m}(\cos\psi) \\ \pm \sin^{m+1}{\psi\over 2}\cos^{m} {\psi\over 2}
P^{\left(m, m+1\right)}_{l-m}(\cos\psi) \end{array}\right),
\end{equation}
where
\begin{equation}
l,m\in \mathbb{Z}, \quad l\ge 0, \quad 0\le m\le l.
\end{equation}
These spinors satisfy
\begin{equation}
\slashed{D}_{S^2} \chi_{l,m}^\pm =\pm i\, a^{-1}\,
\left(l +1\right) 
\chi_{l,m}^\pm\, , \qquad
\not \hskip -4pt D_{S^2} \eta_{l,m}^\pm =\pm i\, a^{-1}\,
\left(l +1\right) 
\eta_{l,m}^\pm\, .
\end{equation}
Here $P^{\alpha,\beta}_n(x)$ are the Jacobi Polynomials:
\begin{equation}
P_n^{(\alpha,\beta)}(x) = { (-1)^n\over 2^n \, n!} (1-x)^{-\alpha}
(1+x)^{-\beta} {d^n\over dx^n} \left[ (1-x)^{\alpha+n}
(1+x)^{\beta+n}\right]\, .
\end{equation}
Now the integrated heat kernel on the unquotiented space $\ss2\otimes\ss2$ is given by\footnote{Following \cite{Banerjee:2010qc} we will compute the heat kernel of $-\slashed{D}^2$, which is the square of $i\slashed{D}$, which means we should multiply a factor of ${1\over 2}$, but we are also considering a Dirac fermion, rather than a Majorana fermion, which yields a factor of 2. These two factors cancel.}
\begin{equation}\label{fs2s2}
K^f_{\ss2\otimes\ss2}\l(a_1,a_2\r)=\l[\sum_{\tilde{\ell}=0}^\infty 4\l(\tilde{\ell}+1\r)e^{-{t\over a_1^2}\l(\tilde{\ell}+1\r)^2}\r]\cdot\l[\sum_{\ell=0}^\infty 4\l(\ell+1\r)e^{-{t\over a_2^2}\l(\ell+1\r)^2}\r].
\end{equation}
Each of the two products corresponds to
\begin{equation}
K^f_{\ss2}\l(a\r)=\sum_{\ell=0}^\infty 4\l(\ell+1\r)e^{-{t\over a^2}\l(\ell+1\r)^2}=\sum_{\ell=0}^\infty 2\sum_{m=-\ell-{1\over 2}}^{\ell+{1\over 2}}e^{-{t\over a^2}\l(\ell+1\r)^2},
\end{equation}
from which we obtain
\begin{equation}
K^f_{S^2\otimes S^2}\l(a_1,a_2\r)= \sum_{\ell'=0}^\infty 2\sum_{m=-\ell'-{1\over 2}}^{\ell'+{1\over 2}}e^{-{t\over a_1^2}\l(\ell'+1\r)^2}\cdot \sum_{\ell=0}^\infty 2\sum_{n=-\ell-{1\over 2}}^{\ell+{1\over 2}}e^{-{t\over a_2^2}\l(\ell+1\r)^2}.
\end{equation}
The modes invariant under the orbifold action \ref{orbact} are still given by the quantization condition
\begin{equation}
m-n=Np, p\in\mathbb{Z},
\end{equation}
where $n$ is the azimuthal quantum number on the sphere with radius $a_2$. 
On imposing this projection as for the scalar field, we find that the integrated heat kernel is given by
\begin{equation}\label{fs2s2n}
K^f_{\s2s2n}= \sum_{\ell'=0}^\infty\sum_{\ell=0}^\infty D^f\l(\ell',\ell\r)e^{-{t\over a_1^2}\l(\ell'+1\r)^2} e^{-{t\over a_2^2}\l(\ell+1\r)^2},
\end{equation}
where 
\begin{equation}
D^f\l(\ell',\ell\r)={4\over N}\sum_{s=0}^{N-1}\chi_{\ell'+{1\over 2}}\l({\pi s\over N}\r)\chi_{\ell+{1\over 2}}\l({\pi s\over N}\r).
\end{equation}
We can separate out the $s=0$ contribution as for the scalar, and write
\begin{equation}
K^{f}_{\s2s2n}\l(a_1,a_2\r)={1\over N}K^f_{\ss2\otimes\ss2}+{4\over N}\sum_{s=1}^{N-1}\sum_{\ell'=0}^\infty\sum_{\ell=0}^\infty\chi_{\ell'+{1\over 2}}\l({\pi s\over N}\r)\chi_{\ell+{1\over 2}}\l({\pi s\over N}\r)e^{-{t\over a_1^2}\l(\ell'+1\r)^2} e^{-{t\over a_2^2}\l(\ell+1\r)^2}.
\end{equation}
The first term on the right-hand side represents the contribution of the smooth part of the quotient space, while the second term may be regarded as the contribution of the fixed points. This form of the answer is suitable for analytic continuation to $\ads2s2n$. However, we do not carry out the analytic continuation here.
\subsection{A Group--Theoretic Interpretation of These Results}
$\adss$ and $\ss2$ are the simplest examples of the so-called symmetric spaces. These are coset spaces $G/H$ where $G$ and $H$ are Lie groups, and $H$ is a subgroup of $G$. Harmonic analysis on such spaces has a group theoretic structure which has been exploited to evaluate the traced integrated heat kernel, on quotients of symmetric spaces where the quotient group acts freely on the sphere or hyperboloid \cite{David:2009xg,Gopakumar:2011qs}. In this section, we draw an interesting parallel between the results obtained previously for freely acting quotient groups and the results obtained here. This parallel also encourages us to expect that the heat kernel may be explicitly evaluated on these conical spaces for arbitrary-spin particles as well, in particular, in higher-dimensional spaces. To draw the parallel, we begin with a brief review of the results of \cite{David:2009xg,Gopakumar:2011qs}. Consider a symmetric space\footnote{Strictly speaking, most of the results of \cite{David:2009xg,Gopakumar:2011qs} 
would hold for quotients of homogenous spaces as well. The additional requirement that the space begin quotiented be symmetric as well leads to a few technical simplifications, and minor simplifications in the final results.} $G/H$ which is quotiented by a discrete group $\Gamma\subset G$, which acts on $G$ from the left. Further, the action of $\Gamma$ is such that it has no fixed points in $G$. In that case, the heat kernel on $\Gamma\backslash G/H$ may be expressed in terms of the heat kernel on $G/H$ using the method of images. For a field transforming in a representation $S$ of $H$, the integrated heat kernel for the Laplacian on the quotient space was found to be
\begin{equation}\label{kerimgs}
K^{S}_{\Gamma}= v\cdot\sum_R \sum_{\g\in \Gamma}\chi_R\l(\g\r)e^{-t E_R^S},
\end{equation}
where $v$ is a volume factor corresponding to the ratio of the volume of $\G\backslash G/H$ to the volume of $G/H$, $R$ are the representations of $G$ which contain $S$ when restricted to $H$, and $E_R^S$ is the eigenvalue of the Laplacian. To obtain this result, $G$ and $H$ were assumed to be compact, and the extension to non-compact spaces $SO\l(N,1\r)/SO(N)$ was carried out by analytic continuation. Though the action of $\zn$ which we have considered on $\ss2$, and the product space $\ss2\otimes\ss2$ has fixed points, the final forms of the answer thus obtained in \eqref{heatksn}, \eqref{scalarkernel}, and \eqref{fs2s2n}, have remarkable similarities to \eqref{kerimgs}. In particular, the values of $\ell$ being summed over in these expressions are precisely the ones determined by the branching rules as in \eqref{kerimgs}. This is possibly related to the fact that for the scalar heat kernel over the cone in flat space, the method of images does in fact produce the correct answer for integer $N$ \cite{
sommer3}. It would be very interesting to make this connection explicit as it would provide a very explicit solution to the heat kernel about these quotient spaces for arbitrary-spin particles. Finally, it was observed in \cite{David:2009xg,Gopakumar:2011qs} that the answer \eqref{kerimgs} could be extended to the non-compact case of the hyperboloids by replacing the Weyl character which appears in \eqref{kerimgs} by the Global (or Harish-Chandra) character. It is natural to ask if the same extension is possible in this case. Though we do not fully explore this question here, we provide some preliminary evidence for this in the context of the $\mathbb{Z}_2$ orbifold of $\adss$. Firstly, we observe that the heat kernel for the scalar on $\ss2/\mathbb{Z}_2$ may also be expressed, using the projection methods of Section \ref{scalarprojectionmethod} as
\begin{equation}\label{heatks2char}
K^s_{\ss2/\mathbb{Z}_2}={1\over 2}\sum_{s=0}^1\sum_{\ell=0}^\infty \chi_{\ell}\l(\pi s\over 2\r)e^{-{t\over a^2}\ell\l(\ell+1\r)}
\end{equation}
We will now show that replacing the Weyl character of $SU(2)$ in \eqref{heatks2char}, and multiplying by a factor of ${1\over 2}$ to account for the relative number of fixed points in $\sn$ and $\adsn$, leads to the asymptotic expansion found in \eqref{scalarads2n} for $N=2$. For this, we require the Harish-Chandra character of the group $SO(2,1)\simeq SL(2,R)$. This is given for the scalar\footnote{The corresponding expression for the fermion is also available in \cite{Lang}.}, by \cite{Lang}
\begin{equation}
d\mu=\frac{1}{2}\frac{\cosh\left(\pi-2\theta\right)\lambda}{\cosh\left(\pi\lambda\right)}d\lambda,
\end{equation}
where $\theta={\pi s\over N}$. For us, $N=2$ and $s=0,1$. The $s=0$ term corresponds to the contribution from the smooth part of $\adsn$, which is just ${1\over N}$ times the heat kernel on $\adss$. We will concentrate on the $s=1$ term, which is responsible for the contribution from the conical singularities. We therefore find
\begin{equation}
K^{\text{conical}}= {1\over 2}\cdot{1\over 2}\int_{0}^\infty d\lambda \frac{1}{\cosh\left(\pi\lambda\right)}e^{-{t\over a^2}\l(\lambda^2+{1\over 4}\r)}\simeq {1\over 8} -{t\over 16 a^2}+\ldots.
\end{equation}
These are precisely the conical terms found in \eqref{scalarads2n} on setting $N=2$ there, which verifies our conjecture. Thus, many of the central results of \cite{David:2009xg,Gopakumar:2011qs} seem to carry over to these quotient spaces as well, leading us to expect that the heat kernel can be explicitly solved for arbitrary rank tensors, on arbitrary-dimensional spheres and hyperboloids in these $\zn$ orbifolds as well. It would be interesting to explore this question further.
\section{The One-loop determinants in the Graviphoton Background}\label{graviphoton}
In this section we will carry out the computation proposed at the end of Section \ref{macro}. In particular, we will consider the $\zn$ orbifold \eqref{orbact} of the near-horizon geometry of a ${1\over 4}$--BPS black hole and compute the log term in the one-loop partition function of an $\cn=4$ vector multiplet. We find that it vanishes, which is consistent with the microscopic results. We now give an overview of our strategy for this computation. Firstly, we note that the spectrum of the quadratic operator for the $\cn=4$ vector multiplet in the attractor geometry has been completely solved for in \cite{Banerjee:2010qc}. This four-dimensional background is $\adss\otimes\ss2$ with graviphoton fluxes running through the $\adss$ and $\ss2$ submanifolds. We refer the reader to \cite{Sen:2007qy} and Section 4.2 of \cite{Banerjee:2010qc} for details. Hence, to determine the spectrum, we can separately analyse the quadratic operators on $\adss$ and $\ss2$. The following is a summary of the results of \cite{Banerjee:2010qc} which we will need for our analysis.

We will first consider the kinetic operator for bosons and then the fermionic kinetic operator. The $\cn=4$ vector multiplet has a single vector field $\mathcal{A}$, two Dirac fermions $\Psi_{1,2}$ and 6 real scalars $\phi_a$, where $1\leq a\leq 6$. Of these, the four scalars $\phi_a$ where $3\leq a\leq 6$ do not mix with any other field. Their action is that of a real scalar field minimally coupled to background gravity. Their contribution to the heat kernel is therefore known from the analysis of Section \ref{lapkernelads2s2n}. The field $\phi_1$ mixes with the transverse modes of $\mathcal{A}$ along $\adss$ on account of the graviphoton flux. The relevant kinetic operator is \cite{Banerjee:2010qc}
\begin{equation}\label{gravads}
-{1\over 2} \int\sqrt{\det g} \l[\l(\begin{array}{cc} \phi_1 & \AAA_{\alpha}\end{array}\r)\l(\begin{array}{cc} -\square- {2\over a^2} & {2i\over a}\varepsilon^{\g\beta}D_{\g}\\ {2i\over a}\varepsilon^{\alpha\g}D_\g & -g^{\alpha\beta}\square+R^{\alpha\beta} +D^{\alpha}D^{\beta} \end{array}\r)\l(\begin{array}{c} \phi_1\\ \AAA_{\beta}\end{array}\r)-\AAA_{\alpha}D^{\alpha}D^{\beta}\AAA_{\beta}\r],
\end{equation}
where the last term is the gauge fixing term. The consequence of this mixing is that the eigenvalues\footnote{We remind the reader that these are the eigenvalues of both the standard Laplacian over scalar fields as well as the Hodge Laplacian over vector fields in $\adss$.} of the kinetic operator shift from ${1\over a^2}\l(\lambda^2+{1\over 4}\r)$ to
\begin{equation}\label{shiftedads}
{1\over a^2}\l[\l(\lambda\pm i\r)^2+{1\over 4}\r].
\end{equation}
The longitudinal and zero modes modes of $\mathcal{A}$ on $\adss$ do not mix with the scalar, and their contribution to the partition function is again known from the analysis of Section \ref{lapkernelads2s2n}. The field $\phi_2$ on the other hand mixes with transverse modes of the vector field $\mathcal{A}$ along the $\ss2$ direction again on account of the graviphoton flux, and the relevant kinetic operator is \cite{Banerjee:2010qc}
\begin{equation}\label{gravs}
-{1\over 2} \int\sqrt{\det g} \l[\l(\begin{array}{cc} \phi_2 & \AAA_{\alpha}\end{array}\r)\l(\begin{array}{cc} -\square+ {2\over a^2} & -{2\over a}\varepsilon^{\g\beta}D_{\g}\\ -{2\over a}\varepsilon^{\alpha\g}D_\g & -g^{\alpha\beta}\square+R^{\alpha\beta} +D^{\alpha}D^{\beta} \end{array}\r)\l(\begin{array}{c} \phi_2\\ \AAA_{\beta}\end{array}\r)-\AAA_{\alpha}D^{\alpha}D^{\beta}\AAA_{\beta}\r],
\end{equation}
where the last term is again the gauge fixing term. In this case, for modes of $\mathcal{A}^T$ and $\phi_{2}$ labelled by $\ell\geq 1$ the eigenvalues shift from ${1\over a^2}\ell\l(\ell+1\r)$ to
\begin{equation}\label{shifteds}
{1\over a^2}\ell\l(\ell-1\r),\quad {1\over a^2}\l(\ell+1\r)\l(\ell+2\r).
\end{equation}
The $\ell=0$ mode of $\phi_2$ does not mix, but its eigenvalue does get shifted from zero to $2\over a^2$. Again, the modes $\mathcal{A}^L$ do not mix with the scalar. In both the $\adss$ and $\ss2$ cases, even though the eigenvalues of the quadratic operator shift from those of the Laplacian, the degeneracies do not. That is to say, that though the eigenvalue shifts from $E_{old}$ to $E_{new}$, the degeneracy of $E_{new}$ is equal to the degeneracy of $E_{old}$. Further, on comparing the kinetic operators in \eqref{gravads} and \eqref{gravs}, it is apparent that they are related to each other by the same analytic continuation as the $\adss$ and $\ss2$ Laplacians are to each other. Finally, the shifted eigenvalues \eqref{shiftedads} and \eqref{shifteds} are related to each other by the same analytic continuation
\begin{equation}
a\mapsto ia,\quad \ell\mapsto i\lambda -{1\over 2},
\end{equation}
as those of the Laplacian on $\adss$ and $\ss2$ \cite{Camporesi:1994ga}. The analysis for the fermions is similar. While the degeneracies do not change, the eigenvalues change from $\pm ia^{-1}\l(\ell+1\r)$ on $\ss2$ to $\pm ia^{-1}\l(\ell+1\pm{1\over 2}\r)$, and from $\pm ia^{-1}\lambda$ on $\adss$ to $\pm ia^{-1}\l(\lambda\pm {i\over 2}\r)$, which are again related to each other through the analytic continuation for fermions \cite{Camporesi:1995fb}. To compute the heat kernel of the $\cn=4$ vector multiplet in the $\zn$ orbifold of $\adss\otimes\ss2$ with the graviphoton flux, we will therefore employ the same strategy as in Section \ref{lapkernelads2s2n}. We will compute the heat kernel in the $\zn$ orbifold of $\ss2\otimes\ss2$, where the two $\ss2$s have radii $a_1$ and $a_2$ respectively, and the mixing along each $\ss2$ is as per \eqref{gravs} and \eqref{shifteds}. We will then analytically continue the answer thus arrived at via
\begin{equation}
a_1\mapsto ia,\quad a_2\mapsto a
\end{equation}
to arrive at the answer on $\ads2s2n$.
\subsection{The Graviphoton Background on $\s2s2n$}
In this section we will compute the heat kernel for an $\cn=4$ vector multiplet in a $\zn$ orbifold of an $\ss2\otimes\ss2$ background with a graviphoton flux running through both $\ss2$s. The scalar $\phi_1$ couples to the gauge field $\mathcal{A}$ on the $\ss2$ of radius $a_1$ via \eqref{gravs}, while the scalar $\phi_2$ is identically coupled to $\mathcal{A}$ on the $\ss2$ of radius $a_2$. The effect of this coupling on the spectrum has already been reviewed above.
\subsubsection{The Bosonic Determinants}
We consider the first the boson contribution which is given by
\begin{equation}\begin{split}
K^b\l(a_1,a_2\r)=4&K^{(s,s)}_{\s2s2n}\l(a_1,a_2\r)+ K^{(s,v+s)}_{\s2s2n}\l(a_1,a_2\r)+
K^{(v+s,s)}_{\s2s2n}\l(a_1,a_2\r)\\&- 2K^{(s,s)}_{\s2s2n}\l(a_1,a_2\r).\end{split}
\end{equation}
Here $K^{(v+s,s)}_{\s2s2n}$ denotes the heat kernel of the vector--scalar mixed system on the $\ss2$ of radius $a_1$ multiplied with the heat kernel for the scalar on the $\ss2$ radius $a_2$, and the subscript means that we have to pick out the $\mathbb{Z}_N$ invariant states and compute the heat kernel over them only. The last term represents the contribution of the ghost fields.
We have already computed $K^{(s,s)}_{\s2s2n}$ in Section \ref{scalarprojectionmethod}. We shall now compute $K^{(v+s,s)}_{\s2s2n}$. The computation for $K^{(s,v+s)}_{\s2s2n}$ is just related by the replacement $a_1\leftrightarrow a_2$. With the spectrum \eqref{gravs}, the heat kernel $K^{(v+s,s)}$ is given by
\begin{equation}\label{s2mixing1}
\sum_{\ell'=0}^\infty\left[\sum_{\ell=1}^\infty D^s(\ell,\ell')e^{-{t\over a_1^2}E_{\ell-1}}+\sum_{\ell=0}^\infty D^s(\ell,\ell')e^{-{t\over a_1^2} E_{\ell+1}}+\sum_{\ell=1}^\infty D^s(\ell,\ell')e^{-{t\over a_1^2}E_\ell}\right]e^{-t\frac{\ell'(\ell'+1)}{a_2^2}},
\end{equation}
where 
\begin{equation}
D^s(\ell,\ell')={1\over N}\sum_{m=0}^{N-1}\frac{\sin\left[\frac{\pi (2\ell+1) m}{N}\right]}{\sin\left[\frac{\pi m}{N}\right]}\frac{\sin\left[\frac{\pi (2\ell'+1) m}{N}\right]}{\sin\left[\frac{\pi m}{N}\right]}\equiv {1\over N}\sum_{m=0}^{N-1}\chi_{\l(\ell,\ell'\r)}\l({\pi m\over N}\r),
\end{equation}
and $E_\ell=\ell\l(\ell+1\r)$. 
By using standard trigonometric identities one can reduce this to the form
\begin{equation}\label{s2mixing3}
K^{(v+s,s)}_{\s2s2n}=3K^s_{\s2s2n}-{4\over N}\sum_{m=1}^{N-1}\sum_{\ell,\ell'=0}^\infty\chi_{\l(\ell,\ell'\r)}\l({\pi m\over N}\r)\sin^2{\pi m\over N}e^{-{t\over a_1^2}E_\ell}e^{-{t\over a_2^2}E_{\ell'}}.
\end{equation}
Putting everything together, we find that
\begin{equation}
K^{b}_{\s2s2n}=8K^{s}_{\s2s2n}-{8\over N}\sum_{m=1}^{N-1}\sum_{\ell,\ell'=0}^\infty\chi_{\l(\ell,\ell'\r)}\l({\pi m\over N}\r)\sin^2{\pi m\over N}e^{-{t\over a_1^2}E_\ell}e^{-{t\over a_2^2}E_{\ell'}}.
\end{equation}
Finally, on using \eqref{scalars2s2n}, we obtain as our final result
\begin{equation}\label{s2mixingfinal}
K^{b}_{\s2s2n}={8\over N}K^{s}_{\ss2\otimes\ss2}+{8\over N}\sum_{m=1}^{N-1}\sum_{\ell,\ell'=0}^\infty\chi_{\l(\ell,\ell'\r)}\l({\pi m\over N}\r)\cos^2{\pi m\over N}e^{-{t\over a_1^2}E_{\ell}}e^{-{t\over a_2^2}E_{\ell'}}.
\end{equation}
\subsubsection{The Fermionic Determinants}
As mentioned previously, the effect of the graviphoton background for fermions on $\ss2$ is to shift the eigenvalues by of the Dirac operator from ${\pm ia^{-1}}\l(\ell+1\r)$ to ${\pm ia^{-1}}\l(\ell+1\pm{1\over 2}\r)$. The degeneracies do not change. As a result, the fermion heat kernel on $\s2s2n$ changes to
\begin{equation}
K=-{1\over N}\sum_{m=0}^{N-1}\sum_{\ell'=0}^\infty\sum_{\ell=0}^\infty\chi_{\l(\ell'+{1\over 2},\ell+{1\over 2}\r)}\l({\pi m\over N}\r)\l(e^{-{t\over a_1^2}\l(\ell'+{3\over 2}\r)^2}+e^{-{t\over a_1^2}\l(\ell'+{1\over 2}\r)^2}\r)\l(e^{-{t\over a_2^2}\l(\ell+{3\over 2}\r)^2}+e^{-{t\over a_2^2}\l(\ell+{1\over 2}\r)^2}\r).
\end{equation}
Again, using standard trigonometric identities we can show that
\begin{equation}
\sum_{\ell=0}^\infty \chi_{\l(\ell+{1\over 2}\r)}\l({\pi s\over N}\r) \l(e^{-{t\over a^2}\l(\ell+{3\over 2}\r)^2}+e^{-{t\over a^2}\l(\ell+{1\over 2}\r)^2}\r)=2e^{-{t\over 4a^2}}\sum_{\ell=0}^\infty \chi_{\ell}\l({\pi s\over N}\r)\cos\l({\pi s \over N}\r) e^{-{t\over a^2}\ell\l(\ell+1\r)}.
\end{equation}
We therefore find
\begin{equation}
K=-{4\over N}e^{-{t\over 4a_1^2}}e^{-{t\over 4a_2^2}}\sum_{s=m}^{N-1}\sum_{\ell=0}^\infty\sum_{\ell'=0}^\infty \chi_{\l(\ell,\ell'\r)}\l({\pi m\over N}\r)\cos^2\l({\pi m \over N}\r) e^{-{t\over a_1^2}E_{\ell'}}e^{-{t\over a_2^2}E_\ell},
\end{equation}
and separating out the $s=0$ term, we find
\begin{equation}
K=-{4\over N}e^{-{t\over 4a_1^2}}e^{-{t\over 4a_2^2}}\l[K^s_{\ss2\otimes\ss2}+\sum_{m=1}^{N-1}\sum_{\ell=0}^\infty\sum_{\ell'=0}^\infty \chi_{\l(\ell',\ell\r)}\l({\pi m\over N}\r)\cos^2\l({\pi m \over N}\r) e^{-{t\over a_1^2}E_{\ell'}}e^{-{t\over a_2^2}E_\ell}\r].
\end{equation}
As there are two Dirac fermions in an $\mathcal{N}=4$ multiplet the overall fermion contribution is given by
\begin{equation}\label{fermionfinalcpt}
K^f=-e^{-{t\over 4a_1^2}}e^{-{t\over 4a_2^2}}K^{b}_{\s2s2n}\l(a_1,a_2\r),
\end{equation}
where we have used \refb{s2mixingfinal}.
\subsection{The Analytic Continuation to $\ads2s2n$}
We will now focus on how the above results may be analytically continued to the heat kernel of the $\cn=4$ vector multiplet in the near horizon geometry of a quarter-BPS black hole. To carry out the analytic continuation, we will use the following prescription. We will take the heat kernels computed on $\s2s2n$ above and continue $a_1\mapsto ia$ and $a_2\mapsto a$ as in Section \ref{lapkernelads2s2n}. We will multiply the resulting functional form of the heat kernel by an overall factor of half\footnote{The origin of the factor of half is as follows. Firstly, for the contribution from the conical singularities it accounts for the fact that the number of fixed points on the AdS quotient is half the number of fixed points on the sphere quotient. Secondly, for the contribution from the smooth part of the manifold, the integrated heat kernel is given by the coincident heat kernel times the volume. The two coincident heat kernels are related by $a\mapsto ia$  \cite{Banerjee:2010qc}. In the analytic continuation, 
the volume of the $\ss2$, which is $4\pi a_1^2$ is replaced by the (regularised) volume of $\adss$, which is $-2\pi a^2$. Hence the overall factor of half, with the continuation $a\mapsto ia$.}. This will yield the regulated heat kernel of the kinetic operator with the graviphoton flux on $\ads2s2n$. Further, we have to be careful to add the $\ell=0$ modes mentioned in Section \ref{hodgelap} when we analytically continue the vector heat kernel from $\ss2$ to $\adss$. We begin with the bosonic contribution to the heat kernel. Following \cite{Banerjee:2010qc}, we have to compute
\begin{equation}\label{bos1}
K^b=4K^{(s,s)}_{\ads2s2n}+K^{(s,v+s)}_{\ads2s2n}+
K^{(v+s,s)}_{\ads2s2n}-2K^{(s,s)}_{\ads2s2n},
\end{equation}
It should be clear that the analytic continuation for most of the terms is straight forward. In particular, $K^{(s,s)}$ gets continued to the corresponding scalar heat kernel in the non-compact geometry and $K^{(s,v+s)}$ gets continued again as before because the field for which the analytic continuation being done is just the scalar, for which there are no subtleties of zero modes. Hence
\begin{equation}\label{bos2}
K^{(s,s)}_{\ads2s2n}(a)={1\over 2}K^{(s,s)}_{\s2s2n}(ia,a), \quad K^{(s,v+s)}_{\ads2s2n}(a)= {1\over 2}K^{(s,v+s)}_{\s2s2n}(ia,a).
\end{equation}
It now remains to compute $K^{(v+s,s)}_{\ads2s2n}$. This may be decomposed as
\begin{equation}\label{vs0}
K^{(v+s,s)}_{\ads2s2n}\l(a\r)=K^{(v_T+v_L+s,s)}_{\ads2s2n}\l(a\r)+K^{(v_0,s)'}_{\ads2s2n}\l(a\r),
\end{equation}
since the zero modes of the gauge field do not mix with the scalar field $\phi_1$ on $\adss$. The prime on the superscript of the second term on the right hand side above reminds us that this is the heat kernel over the non-zero modes obtained by tensoring the zero modes of the vector field on $\adss$ with the non-zero modes of the scalar on $\ss2$. This has already been evaluated in \refb{znonz} and found to be,
\begin{equation}\label{vs1}
K^{(v_0,s)'}_{\ads2s2n}\l(a\r)=-K^s_{\ss2/\mathbb{Z}_N}\l(a\r)+1.
\end{equation}
The first term $K^{(v_T+v_L+s,s)}_{\ads2s2n}\l(a\r)$ is given by
\begin{equation}\label{vs2}
K^{(v_T+v_L+s,s)}_{\ads2s2n}\l(a\r)={1\over 2}\l[K^{(v_T+v_L+s,s)}_{\s2s2n}\l(ia,a\r)+2K^s_{\sn}\l(a\r)\r],
\end{equation}
where the $2K^s_{\sn}$ being added is because when we analytically continue the longitudinal and transverse modes of the vector fields from $\ss2$ to $\adss$ we have to add back the $\ell=0$ modes of the scalars $\phi_1$ and $\phi_2$ as discussed in Section \ref{hodgelap}.
We therefore find from \eqref{vs0}, \eqref{vs1} and \eqref{vs2} that
\begin{equation}\label{bos3}
K^{\l(v+s,s\r)}_{\ads2s2n}\l(a\r)={1\over 2}K^{\l(v+s,s\r)}_{\s2s2n}\l(ia,a\r)+1.
\end{equation}
Finally, putting \eqref{bos1},\eqref{bos2} and \eqref{bos3} together, we find that the bosonic heat kernels contribute
\begin{equation}
K^b = {1\over 2}\l[2K^s_{\s2s2n}\l(ia,a\r)+K^{\l(s,v+s\r)}\l(ia,a\r)+K^{\l(v+s,s\r)}_{\s2s2n}\l(ia,a\r)\r]+1,
\end{equation}
which simplifies to
\begin{equation}
K^b = {1\over 2}K^b_{\s2s2n}\l(ia,a\r)+1.
\end{equation}
Now the fermionic contribution may be analytically continued from the expression \refb{fermionfinalcpt} to obtain
\begin{equation}
K^f=-{1\over 2}K^b_{\s2s2n}\l(ia,a\r).
\end{equation}
Then the contribution from the non-zero modes to the heat kernel is finally
\begin{equation}
K_{\text{non-zero}}=K^{b}+K^f=+1
\end{equation}
Therefore the contribution to the log term from non-zero modes is
\begin{equation}
S^{\text{non-zero}}_{\text{log}}={1\over 2}\int_{\epsilon\over a^2} {d\bar{s}\over\bar{s}} (+1)=\log a.
\end{equation}
We've already counted the number of zero modes on the 4d geometry, which is -1, so
\begin{equation}
S^{\text{zero}}_{\text{log}}=(-1)\log a.
\end{equation}
Thus the two terms, when added, cancel each other and the net logarithmic contribution vanishes.
\begin{equation}
S_{\text{log}}=0.
\end{equation}
\section{Conclusions}
In this paper we computed the heat kernel for a single $\cn=4$ vector multiplet in a $\zn$ orbifold of the near-horizon geometry of a quarter-BPS black hole in $\cn=4$ supergravity. We found that the contribution proportional to $\log a$ vanishes for arbitrary values of $N$, in accordance with the expectation from microstate counting. To carry out this computation, we developed new techniques to compute the short-time asymptotic behaviour of the heat kernel over spheres and hyperboloids with $\zn$ orbifolds. Though we worked mostly with $\ss2$, $\adss$, their product spaces and their $\zn$ orbifolds, we also obtained group-theoretic forms for the heat kernel suggestive of a possible solution for the heat kernel over arbitrary-spin fields in higher-dimensional spheres and hyperboloids. It would firstly be very interesting to make this connection to the results previously obtained for freely acting quotient groups in \cite{David:2009xg,Gopakumar:2011qs}. Finally, the motivation for this work was to concretely 
match the logarithmic corrections about exponentially suppressed saddle points of the quantum entropy function with the corresponding microscopic answers for $\cn=4$ and $\cn=8$ string theory along the lines of \cite{Banerjee:2010qc,Banerjee:2011jp} for the leading saddle-point. This requires us to extend the analysis of this paper to the gravity multiplet as well, which is work in progress. In addition, these methods would also be useful in the case of extremal black holes in AdS$_4$ constructed in \cite{Cacciatori:2009iz,Hristov:2010ri, Dall'Agata:2010gj}. In this case the $AdS_2$ and $S^2$ have different radii which in our expressions can be obtained by analytically continuing one of the $\ss2$s to an $\adss$ of radius different than the other $\ss2$.
\acknowledgments
We would like to thank Pallab Basu, Archisman Ghosh, Chethan Krishnan, Sameer Murthy, Suvrat Raju, and Spenta Wadia for helpful discussions. We would especially like to thank Shamik Banerjee, Justin David, Rajesh Gopakumar Soo-Jong Rey, and Ashoke Sen for several discussions, comments, and correspondence. SL thanks HRI, Allahabad, NORDITA, Stockholm, NIKHEF Amsterdam, and the GGI, Florence's program ``Higher Spins, Strings and Dualities'' for hospitality while part of this work was carried out. SL's work is supported by National Research Foundation of Korea grants 2005-0093843, 2010-220-C00003 and 2012K2A1A9055280. During the initial stages of this work, SL was supported by a postdoctoral fellowship at the ICTS--TIFR.
\appendix
\section*{Appendix}
\section{The Scalar Laplacian on $\ss2$ and the Analytic Continuation to $\adss$}\label{A}
As a zeroth order check of our method of evaluating the heat kernel expansion by explicitly enumerating the eigenvalues and degeneracies of the Laplacian and using the Euler-Maclaurin formula, we will show that this precisely reproduces results for the integrated heat kernel arrived at previously in \cite{Banerjee:2010qc}. Additionally, we will recover the result for the contribution of a single massless minimally coupled scalar to the logarithmic correction to the entropy of an extremal black hole also obtained in \cite{Banerjee:2010qc}. 
We start with the expression for the integrated heat kernel on the two-sphere
\begin{equation}
S_g=\sum_{\ell=0}^\infty\l(2\ell+1\r)e^{-t{\ell(\ell+1)\over a^2}}.
\end{equation}
Using the Euler-Maclaurin formula \eqref{eulermac} this can be expressed as
\begin{equation}
S_g=={1\over 2}f(0)+\int_0^\infty d\ell \l(2\ell+1\r)e^{-t{\ell(\ell+1)\over a^2}}+\sum_{k=2}^\infty {B_k\over k!}\l(f^{(k-1)}(\infty)-f^{(k-1)}(0)\r),
\end{equation}
where
\begin{equation}
f(x)\equiv \l(2x+1\r)e^{-t{x(x+1)\over a^2}}.
\end{equation}
We will now extract the short-time asymptotic behaviour of $S_g$ from the above expression. Firstly, note that the integral over $\ell$ can be done to obtain
\begin{equation}
S_g={1\over 2}+{a^2\over t}-\sum_{k=2}^\infty {B_k\over k!}f^{(k-1)}(0),
\end{equation}
where we have also used the fact that $f$ and all its derivatives vanish at $\infty$. Also, to extract the log term from the heat kernel, we need the order 1 term in the $t$ expansion of the scalar heat kernel on $\adss\otimes\ss2$. The scalar heat kernel on this product space is just the product of the scalar heat kernel on $\adss$ times the scalar heat kernel on $\ss2$. We know that the heat kernels go as ${1\over t}$ so we will neglect terms from $\mathcal{O}(t^2)$ onwards in this expansion. Expanded out to any finite power of $t$, the function $f$ will be polynomial in $x$ and therefore the sum over $k$ above is a finite sum and can be evaluated completely. Upon doing so, we find
\begin{equation}\label{A1s2}
S_g={1\over 3}+{a^2\over  t}+{t\over 15 a^2}+\mathcal{O}(t^2).
\end{equation}
Now we would like to analytically continue this to the Anti-de Sitter case. Note that this is the integrated heat kernel, and we cannot naively continue this, as there is a volume divergence in AdS which will be missed. To do this continuation note that
\begin{equation}
S_g=\sum_{\ell=0}^\infty\l(2\ell+1\r)e^{-t{\ell(\ell+1)\over a^2}}=(\text{Vol.}_{S^2})\sum_{\ell=0}^\infty{2\ell+1\over 4\pi a^2}e^{-t{\ell(\ell+1)\over a^2}}.
\end{equation}
The last sum is the coincident heat kernel, which can be safely analytically continued via $a\mapsto ia$. The integrated heat kernel on AdS can hence be written down by analytic coninuation as
\begin{equation}
A_g=(\text{Vol.}_{AdS_2})\l(-{1\over 12\pi a^2}+{1\over 4\pi t}+{t\over 60\pi a^4}+\mathcal{O}(t^2)\r).
\end{equation}
The expression for the coincident heat kernel above matches precisely with the series expansion in $\bar{s}={t\over a^2}$ of (2.15) of \cite{Banerjee:2010qc}. Then the integrated heat kernel on $AdS_2\times S^2$ is given by
\begin{equation}
K(t)=-2\pi a^2\l(-{1\over 12\pi a^2}+{1\over 4\pi t}+{t\over 60\pi a^4}+\mathcal{O}(t^2)\r)\cdot\l({1\over 3}+{a^2\over  t}+{t\over 15 a^2}+\mathcal{O}(t^2)\r),
\end{equation}
where $-2\pi a^2$ is the regularized volume of $\adss$ as obtained in \eqref{regvol}. 
Then, the one-loop effective action is given by
\begin{equation}
\delta S= {1\over 2}\int_\epsilon^\infty{dt\over t}K(t),
\end{equation}
from which the one-loop correction to black hole entropy may be extracted
\begin{equation}
\delta S_{BH}=\int_{\epsilon\over a^2}^\infty{d\bar{s}\over \bar{s}}\l({1\over 12}-{1\over 4 \bar{s}}+{\bar{s}\over 60}+\mathcal{O}\l(\bar{s}^2\r)\r)\cdot\l({1\over 3}+{1\over\bar{s}}+{\bar{s}\over 15}+\mathcal{O}\l(\bar{s}^2\r)\r).
\end{equation}
The log contribution, as before, comes from the order 1 term in the integrand, given by ${1\over 180 \pi a^2}$. We therefore obtain
\begin{equation}
\delta S_{BH}=-{1\over 180}\log{a^2\over \epsilon}.
\end{equation}
This matches with the expression found in \cite{Banerjee:2010qc}.
\section{Anti-periodic Fermions on $\sn$ and $\adsn$}
In this appendix, we will compute the heat kernel for anti-periodic fermions on $\sn$ and $\adsn$. Unlike scalars and spin-1 fields, these modes are not a subset of the anti-periodic modes on $\ss2$ and $\adss$. While these are not relevant to the computation of the one-loop determinants in orbifolds of the graviphoton background considered in Section \ref{graviphoton}, we shall briefly outline the main elements of the analysis for these fermions as well. The main result which we shall use is that though these modes are not a subset of the anti-periodic modes found in \cite{Camporesi:1995fb}, the same analytic continuation as found in \cite{Camporesi:1995fb} for the unorbifolded case can be used to relate the spectrum of the Dirac operator on $\sn$ and $\adsn$ as well. It is straighforward to follow through the analysis of \cite{Camporesi:1995fb} to verify this statement. We can therefore use the Euler-Maclaurin formula to compute the integrated heat kernel on $\sn$ and further prescribe an analytic 
continuation to obtain the integrated heat kernel on $\adsn$. The last expression is, to our knowledge, new. We will first consider the calculation on $\sn$. The spectrum of the Dirac operator is given by \cite{Fursaev:1996uz}
\begin{equation}
\lambda=\pm\l(N\l(p+{1\over 2}\r)+q+{1\over 2}\r), \quad p,q\in\mathbb{Z}_+,
\end{equation}
and the degeneracy of each eigenvalue is 2. Then the square of the Dirac operator has eigenvalues 
\begin{equation}
\lambda=\l(N\l(p+{1\over 2}\r)+q+{1\over 2}\r)^2, \quad p,q\in\mathbb{Z}_+,
\end{equation}
where each eigenvalue is four-fold degenerate. But there are more degeneracies than this. To see the point, consider an eigenvalue $\ell =Nr+s$, where $0\leq s\leq N-1$. There are $\l(r+1\r)$ ways of realising this as $Np+q$, corresponding to $p=0,1,\ldots,r$. Therefore the heat kernel for the Dirac operator can be given by\footnote{Following \cite{Banerjee:2010qc} we will compute the heat kernel of $-\slashed{D}^2$, which is the square of $i\slashed{D}$, which means we should multiply a factor of ${1\over 2}$, but we are also considering a Dirac fermion, rather than a Majorana fermion, which yields a factor of 2. These two factors cancel.}
\begin{equation}
K^f_{\sn}=-\sum_{r=0}^\infty\sum_{s=0}^{N-1}4\l(r+1\r)\l[e^{-{t\over a^2}{\l(Nr+s+{N+1\over 2}\r)^2}}\r]\equiv -\sum_{s=0}^\infty k_s,
\end{equation}
where
\begin{equation}
k_s=\sum_{r=0}^\infty 4\l(r+1\r)e^{-{t\over a^2}\l(Nr+s+{N+1\over 2}\r)^2}.
\end{equation}
Now, as for the scalar on $\sn$ in Section \ref{scalarsn}, we use the Euler-Maclaurin formula to find that the heat kernel for Dirac fermions on $\sn$ is given by
\begin{equation}
K^f_{\sn}={1\over N}K^f_{\ss2}+\frac{N^2-1}{6 N}+\frac{\left(7 N^2+17\right) \left(N^2-1\right) t}{720 a^2 N}+\mathcal{O}\l(t^2\r).
\end{equation}
This matches with the $t^0$ term computed in \cite{Fursaev:1996uz}. Then, the analytic continuation of this answer to the AdS case is
\begin{equation}
K^f_{\adsn}={1\over N}K^f_{\adss}+{1\over 2}\l[\frac{N^2-1}{6 N}-\frac{\left(7 N^2+17\right) \left(N^2-1\right) t}{720 a^2 N}\r]+\mathcal{O}\l(t^2\r).
\end{equation}
Here, we introduced a factor of half in the contribution from the conical singularity, as $\adsn$ has only one such singularity, not two, as in the case of $\sn$. We also analytically continued via $a\mapsto -ia$ in these terms, as per the rules outlined in the main text.
\section{Zero Modes on $\adss$ by Zeta Function Regularization}\label{zetafunc}
In this section we will evaluate the number of zero modes on $\adss$ and $\ads2s2n$ by a different regularization. On non-compact manifolds like $\adss$ this is typically a divergent quantity which we define as
\begin{equation}
n_0 =\sum_{\ell\in \mathbb{Z}}\int d\theta d\eta\sqrt{g} g^{mn} f^{\ell *}_m f_n^{\ell},
\end{equation}
where 
\begin{equation}
f_m^\ell=\nabla_m\f^\ell,\quad \f^\ell={1\over\sqrt{2\pi\vert\ell\vert}}\l[\sinh\eta\over 1+\cosh\eta\r]^{\vert\ell\vert}e^{i\ell\theta},\, l=\pm 1,\pm 2,\ldots,
\end{equation}
are the discrete zero modes of the vector Laplacian on $\adss$. We will first evaluate this answer using the methods of \cite{Banerjee:2010qc}, and then show how a zeta function regularisation gives rise to the same result. We begin with the method of \cite{Banerjee:2010qc}. In this case, we note that $\adss$ is a homogeneous space and we can evaluate the integrand $\sum_\ell g^{mn} f^{\ell *}_m f_n^{\ell}$ at any point on the space, in particular, the origin. In that case only the $\ell=\pm 1$ modes contribute. Then the number of zero modes is given by
\begin{equation}
n_0={2\over 4\pi a^2}\l(\text{Vol.}\adss\r)=-1,
\end{equation}
where we have used $-2\pi a^2$ as the regularized volume of $\adss$. Now we will evaluate the above answer by doing the volume integral explicitly. We then find, on doing the $\eta$ integral
\begin{equation}
n_0=\sum_{\ell\in\mathbb{Z}-{0}}^\infty \int_0^{2\pi} d\theta {1\over 2\pi} = 2\sum_{\ell=1}^\infty 1 =2\zeta\l(0\r)=-1.
\end{equation}
Hence the zeta function regularization gives the same answer as obtained by regularizing the volume divergence. Now suppose we put the orbifold $\theta\mapsto\theta+{2\pi\over N}$ on $\adss$. The modes $f^\ell_m$ which survive the projection are the ones for which $\ell= Np$, where $p=\pm 1,\pm 2,\ldots,\infty$. On going through the above procedure, we again find that
\begin{equation}
n_0=-1.
\end{equation}
This matches with the number of zero modes \eqref{nzero} obtained in the main text.

\bibliography{paper}
\bibliographystyle{unsrt}
\end{document}